\documentclass[journal]{IEEEtran}
\usepackage{amsmath,amsfonts}
\usepackage{amssymb}
\usepackage{algorithm}
\usepackage{algorithmicx}
\usepackage{algpseudocode}
\usepackage{array}
\usepackage[caption=false,font=normalsize,labelfont=sf,textfont=sf]{subfig}
\usepackage{textcomp}
\usepackage{stfloats}
\usepackage{color}
\usepackage{url}
\usepackage{cite}
\usepackage{booktabs}
\usepackage{verbatim}
\usepackage{graphicx}
\usepackage{subcaption}
\usepackage{stfloats}
\usepackage[table]{xcolor}  
\usepackage{amsthm}
\theoremstyle{definition}
\newtheorem{proposition}{Proposition}
\newtheorem{remark}{Remark}
\newtheorem{lemma}{Lemma} 
\usepackage{caption}
\def\BibTeX{{\rm B\kern-.05em{\sc i\kern-.025em b}\kern-.08em
    T\kern-.1667em\lower.7ex\hbox{E}\kern-.125emX}}
\usepackage{balance}

\usepackage{etoolbox}
\makeatletter
\patchcmd{\@makefntext}{\@makefnmark}{}{}{}
\makeatother

\begin{document}
\title{Constellation Design and System Analysis for Sensing-assisted Vehicular Communications}
\author{ Jiaming Hu, Kaitao Meng, Member, IEEE, Borui Du, Christos Masouros, \textit{Fellow, IEEE}
\thanks{}}
\maketitle

\footnotetext{\relax J. Hu, B. Du, and C. Masouros are with Department of Electronic and Electrical Engineering, University College London, London, United Kingdom (e-mails: \{jiaming.hu.19, borui.du.24, c.masouros\}@ucl.ac.uk)}

\footnotetext{\relax K. Meng is with the Department of Electrical and Electronic Engineering, University of Manchester, London, United Kingdom (e-mail: kaitao.meng@manchester.ac.uk)}

\begin{abstract}
 This paper investigates constellation design for sensing-assisted vehicular communications with predictive beamforming. We study how constellation-dependent sensing accuracy propagates through vehicle-state prediction and ultimately affects beam alignment and communication reliability in dynamic vehicular ISAC systems. To this end, we develop a unified analytical framework that characterizes the complete error propagation chain from delay–Doppler–angle estimation, to Kalman-based beam prediction, and finally to data symbol detection. Closed-form expressions are derived for parameter-estimation root-mean-square error (RMSE), one-step transmit beam prediction error, two-step receive beam prediction error, beamforming gain degradation, and symbol error rate (SER) in massive multiple-input multiple-output orthogonal frequency division multiplexing systems. The analysis reveals that the fourth moment and inverse second moment of the constellation govern sensing accuracy under matched filtering and reciprocal filtering, respectively, and further determine beam prediction performance and communication reliability. Based on these results, we show that the optimal constellation depends on the prediction-noise regime: sensing-oriented constellations are preferred when prediction uncertainty is high due to their superior sensing accuracy, whereas communication-oriented constellations become preferable when prediction uncertainty is low. To further improve robustness, a deep-learning-assisted Kalman filtering framework and a training-based pre-compensation strategy are proposed to reduce prediction errors and mitigate beam-misalignment-induced distortions. Simulation results validate the analytical derivations and demonstrate the effectiveness of the proposed designs in dynamic vehicular ISAC environments.
\end{abstract}


\section{Introduction}

\IEEEPARstart{I}{n} recent years, Integrated Sensing and Communications (ISAC) has attracted significant attention \cite{kaitao1}\cite{9737357}, owing to its capability to enhance spectral efficiency by enabling simultaneous data transmission and environmental sensing within shared hardware and spectrum resources. For communication-centric ISAC systems, communication signals are not only used for information delivery but also exploited to extract environmental information, supporting applications such as target localisation \cite{10475369}, tracking \cite{10977743}, and predictive beamforming \cite{FanKalman}. These functionalities become increasingly critical in large-scale deployments, such as network-level ISAC \cite{kaitao2, kaitao3}, where accurate state estimation must be maintained in dynamic and uncertain environments.

Vehicular-to-everything (V2X) communications, driven by the rapid development of autonomous driving technologies, have emerged as a key application scenario for ISAC techniques \cite{9779322}. In such systems, vehicles are required to communicate while simultaneously sensing the environment or being sensed by infrastructure, making V2X inherently compatible with ISAC paradigms \cite{11091493, 10007643}. For instance, waveform design has been extensively investigated as a means of shaping the ambiguity function to improve localisation and parameter estimation accuracy \cite{10944644}. In parallel, machine learning-based frameworks have been developed to support the joint optimisation of radio resources, thereby balancing sensing and communication performance in a data-driven manner \cite{10606449}. Reinforcement learning techniques have also been applied to RIS-aided V2X ISAC systems to jointly improve sensing accuracy and communication security \cite{10876793}.

Among the existing ISAC-enabled techniques, sensing-assisted communication has received particular attention for predictive beamforming in high-mobility V2X scenarios. To effectively track a vehicle and construct a communication link, \cite{7914742} uses beam training, which exhaustively transmit and receive pilot signals across all candidate beam directions and select the beam corresponding to the maximum received power. While effective, such methods incur significant latency and overhead, rendering them unsuitable for highly dynamic scenarios. To further optimise the framework, the work in \cite{FanKalman} exploits the Markovian dynamics of vehicular motion and employs Kalman filtering techniques, including the Kalman Filter (KF) and Extended Kalman Filter (EKF) \cite{KalmanFilter, extendedKalmanFilter}, to track vehicle states and enable predictive beam alignment. Specifically, one-step prediction is used for transmit beamforming at the base station, while two-step prediction supports receive beamforming at the vehicle. This framework has been further extended to more complex scenarios, such as curvilinear road geometries \cite{9947033}. Despite these advances, most existing models still rely on simplified assumptions, such as vehicles strictly following predefined trajectories, which implies low prediction noise variance. In practice, however, vehicular motion is affected by complex driving conditions, leading to significant uncertainty and error propagation, particularly in multi-step prediction schemes.

Beyond system-level optimisation, the fundamental design of communication signals provides another important avenue for improving ISAC performance. In particular, constellation design plays a crucial role in balancing sensing and communication objectives \cite{11391499, geiger2025constellation, 11535018}. Prior studies have revealed a deterministic-random trade-off, where Gaussian-like constellations favour communication performance, while constant-modulus constellations are advantageous for sensing \cite{10206462}. Techniques such as probabilistic constellation shaping \cite{10685511}, geometric shaping \cite{10938348}, and their joint optimisation \cite{10880638} have demonstrated strong performance in static ISAC scenarios. Furthermore, machine learning-based end-to-end approaches have shown the capability to learn constellation structures that jointly optimise sensing and communication metrics \cite{alberge2018deep, 10938348}. However, existing constellation design methods predominantly focus on static or quasi-static systems, and their extension to dynamic sensing-assisted communication scenarios remains largely unexplored.

\begin{figure}
    \centering
    \includegraphics[width=\linewidth, trim={0cm 1.2cm 0cm 0.8cm},
    clip]{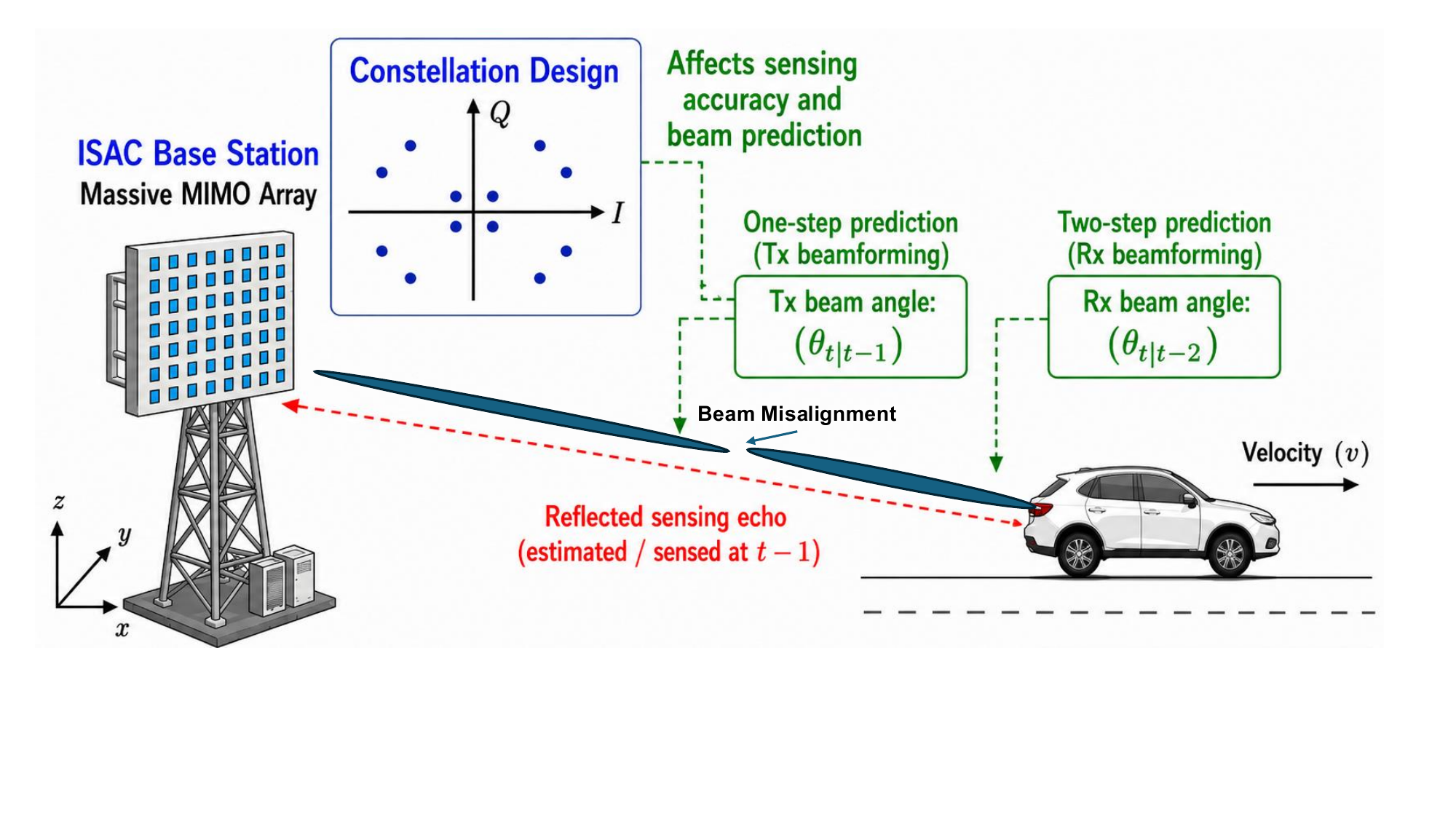}
    \caption{Scenario for Sensing Assisted Communication}
    \label{fig: scenario}
\end{figure}

Motivated by the above limitations, this paper investigates constellation design for dynamic sensing-assisted vehicular communication systems, as shown in Fig.~\ref{fig: scenario}. A key challenge is that prediction uncertainty alters the role of sensing information in beamforming and link maintenance, thereby shifting the preference between sensing- and communication-oriented constellations. It is therefore necessary to examine how constellation choices should adapt to different levels of prediction noise. To this end, we analyse the impact of constellation design on sensing accuracy and communication performance under varying prediction-noise conditions. Closed-form expressions for the root mean square error (RMSE) of parameter estimation and beam direction prediction are derived under a massive MIMO framework which is critical for our framework \cite{FanKalman}. Specifically, we characterise both one-step prediction at the base station and two-step prediction at the vehicle, revealing the impact of prediction error propagation on system performance. Furthermore, to account for practical uncertainty in prediction noise, we evaluate the communication performance using an end-to-end learning-based framework \cite{10938348} across different noise regimes, thereby identifying the sensing-communication trade-off from a constellation design perspective. Based on these insights, we further propose a learning-enhanced Kalman filtering approach to improve tracking accuracy and reduce prediction uncertainty. In addition, we introduce a training-based pre-compensation strategy to mitigate phase distortion induced by system errors, leveraging the generalisation capability of machine learning. These designs enable improved robustness of sensing-assisted communication systems under dynamic conditions. The main contributions of this paper are summarised as follows:
\begin{itemize}
    \item We derive the closed-form expression for the beam direction error under sensing-assisted vehicular communication, explicitly characterising how constellation-dependent statistics, namely the fourth moment and inverse second moment, affect the root-mean-square error of the estimated beam angle. By analysing the equation, the effect of estimation error at the base station onto the predicted angle is numerically explained, which can offer useful insights for constellation shaping in practical sensing-assisted vehicular communication system designs.

    \item We then characterise the exact symbol error rate of the system for different constellations, and to further optimise the system performance, we propose two optimisation strategies including a learning based Kalman filter placed at the transmitter to optimise the trajectory prediction accuracy, and a pre-compensation training strategy focusing on the receiver placed at the vehicle to combat the phase change and energy loss.

\end{itemize}

Beyond the proposed analytical framework and optimization strategy, our study also provides several practical insights into sensing-assisted vehicular communications. First, the preferred constellation design is not fixed but strongly depends on the prediction uncertainty. As sensing noise increases, constellations with larger fourth moments become increasingly beneficial because they reduce beam prediction errors, despite sacrificing conventional communication efficiency. Second, the proposed learning-enhanced kalman filter framework becomes increasingly advantageous under high-uncertainty scenarios, where the existing evolution model cannot well-predict the vehicle movement. These observations reveal a fundamental sensing–communication trade-off and provide practical guidelines for adaptive constellation design in future sensing-assisted wireless systems.

 \textit{{Notation:}} {Scalars are denoted by italic letters, vectors by bold lowercase letters, and matrices by bold uppercase letters. For a generic time-varying quantity at time $t$, $a_t$: $a_t$, $\widehat{a}_t$, and $\widetilde{a}_{t|t'}$ denote the true value, the estimated value obtained through sensing, and the predicted value obtained from evolution model, respectively. Kalman filtered result is denoted as $\widehat{a}_{t|t}$. In particular, $\theta_t$ denotes the true vehicle azimuth angle, $\widehat{\theta}_t$ denotes the estimated vehicle azimuth angle, $\widetilde{\theta}_{t|t-1}$ denotes the one-step predicted azimuth angle used for RSU transmit beamforming, and $\widetilde{\theta}_{t|t-2}$ denotes the two-step predicted azimuth angle used for UE receive beamforming. $\widehat{\theta}_{t|t} $ denotes the Kalman-filtered vehicle azimuth angle at time instance $t$. The superscripts $(\cdot)^{\mathrm{T}}$, $(\cdot)^{\mathrm{H}}$, and $(\cdot)^{-1}$ denote transpose, Hermitian transpose, and matrix inverse, respectively. In addition, $\mathbf{I}$ denotes an identity matrix with appropriate dimensions.
}

\section{System Model}
    \label{sec: System Model}

 {This section presents the system model of the considered sensing-assisted communication system. At time instant $t-2$, the RSU first senses the vehicle and performs Kalman filtering, obtaining the vehicle state estimate $\widehat{\mathbf{x}}_{t-2}$,} where {$\mathbf{x}=[\theta,d,v,\alpha]^{\mathrm{T}}$ denotes the true vehicle state composed of azimuth angle,} propagation distance, {radial} velocity, and complex channel gain. Based on $\widehat{\mathbf{x}}_{t-2}$, the RSU predicts the two-step {UE receive beam direction $\widetilde{\theta}_{t|t-2}$ for time instant $t$. Before time instant $t$, the RSU sends the corresponding UE receive beam index to the vehicular UE through a control link. At time instant $t-1$, the vehicle state estimate is updated by the RSU to $\widehat{\mathbf{x}}_{t-1}$ through Kalman filtering and the one-step RSU transmit beam direction is then predicted. $\widetilde{\theta}_{t|t-1}$ for time instant $t$. At time instant $t$, the RSU transmits the ISAC signal towards $\widetilde{\theta}_{t|t-1}$, and the vehicular UE receives the communication signal using the receive beam indicated by the RSU. The RSU transmit signal model, the RSU received signal model for sensing, and the vehicular UE received signal model for communication are presented in Sections~\ref{sec: RSU Transmitted ISAC Signal Model}, \ref{sec: RSU Received Signal Model for Sensing}, and \ref{sec: Vehicle Received Signal Model for Communication}, respectively. The Kalman filtering framework is introduced in Section~\ref{sec: Evolution and Measurement Model}}.

\subsection{{RSU Transmitted ISAC Signal Model}}
  \label{sec: RSU Transmitted ISAC Signal Model}

{The RSU transmits an OFDM-based ISAC signal. Let $x(t)$ denote the continuous-time transmit signal. Without oversampling, it is given by
}\begin{equation}
{x(t) \triangleq 
\sum\limits_{k=0}^{K-1}
\sum\limits_{m=0}^{M-1}
s_{m,k} u(t-kT)e^{j2\pi m\Delta f(t-kT)},
}\end{equation}  
 {where $K$ is the number of OFDM symbols}, $M$ is the number of subcarriers{, $s_{m,k}$ is the baseband symbol on the $m$th subcarrier of the $k$th OFDM symbol, $u(t)$ is the rectangular pulse, $T$ is the OFDM symbol duration, $\Delta f$ is the subcarrier spacing, and $T_s$ is the sampling interval.
} 

 {A ULA with half-wavelength antenna spacing is considered. The transmit and receive steering vectors are defined as
}\begin{align}
{\mathbf{a}(\theta, N_{\text{T}})
}&{\triangleq
\sqrt{\frac{1}{N_{\text{T}}}}
}[{1, e^{-j\pi \cos(\theta)}, \ldots, e^{-j\pi (N_{\text{T}}-1)\cos(\theta)}}]{^T, }\\
{\mathbf{b}(\theta, N_{\text{R}})
}&{\triangleq
\sqrt{\frac{1}{N_{\text{R}}}}
}[{1, e^{-j\pi \cos(\theta)}, \ldots, e^{-j\pi (N_{\text{R}}-1)\cos(\theta)}}]{^T,
}\end{align} 
 {where $\theta$ is the azimuth angle, and $N_{\text{T}}$ and $N_{\text{R}}$ are the numbers of transmit and receive antennas, respectively.
} 

 {Maximum-ratio transmission (MRT) beamforming is adopted for the transmitter (RSU). Based on the one-step predicted angle $\widetilde{\theta}_{t|t-1}$, the RSU} transmit beamforming vector is {defined as $\mathbf{w}_{\text{T},t}^{\text{RSU}} \triangleq \mathbf{a}(\widetilde{\theta}_{t|t-1}, N_{\text{T}}^{\text{RSU}})$, where $N_{\text{T}}^{\text{RSU}}$ is the number of RSU transmit antennas. The beamformed RSU transmit signal is
}\begin{align}
 {\mathbf{x}_{\text{T},t}^{\text{RSU}}(t)
    =
    \mathbf{w}_{\text{T},t}^{\text{RSU}}
    \sum\limits_{k=0}^{K-1}
    \sum\limits_{m=0}^{M-1}
    s_{m,k} u(t-kT)e^{j2\pi m\Delta f(t-kT)},
}\end{align}
{where $\mathbf{x}_{\text{T},t}^{\text{RSU}}(t)$ is the continuous-time RSU transmit signal vector. After sampling and demodulation, the corresponding subcarrier-domain signal is
}\begin{align}
 {\mathbf{x}_{\text{T},t,m,k}^{\text{RSU}}
    =
    \mathbf{w}_{\text{T},t}^{\text{RSU}}s_{m,k},
}\end{align}
{where $\mathbf{x}_{\text{T},t,m,k}^{\text{RSU}}$ denotes the RSU transmit signal vector on the $m$th subcarrier of the $k$th OFDM symbol.
}

\subsection{{RSU Received Signal Model for Sensing}}
\label{sec: RSU Received Signal Model for Sensing}

  Let $\theta_t$ denote the predicted {vehicle angle at time instant $t$. The RSU} receive and transmit steering vectors {associated with $\theta_t$ are $\mathbf{b}_t^{\text{RSU}} \triangleq \mathbf{b}(\theta_t, N_{\text{R}}^{\text{RSU}})$ and $\mathbf{a}_t^{\text{RSU}} \triangleq \mathbf{a}(\theta_t, N_{\text{T}}^{\text{RSU}})$, respectively. Under maximum-ratio combining, $\mathbf{w}_{\text{R},t}^{\text{RSU}} \triangleq \mathbf{b}_t^{\text{RSU}}$.
}

{Define ${G_{\text{R},t}^{\text{RSU}}} \triangleq |\mathbf{w}_{\text{R},t}^{\text{RSU},H}\mathbf{b}_t^{\text{RSU}}|^2$ and ${G_{\text{T},t}^{\text{RSU}}} \triangleq |\mathbf{a}_t^{\text{RSU},H}\mathbf{w}_{\text{T},t}^{\text{RSU}}|^2$. The RSU received signal for sensing is
}\begin{align}
{\label{eq: y_m,k}
       \mathbf{y}^{\text{RSU}}_{m,k} \triangleq
     \alpha^{\text{RSU}}
     e^{j2\pi \left(f_{\text{D}}^{\text{RSU}}kT - m\tau^{\text{RSU}} \Delta f\right)}
     \sqrt{G_{\text{R},t}^{\text{RSU}}G_{\text{T},t}^{\text{RSU}}}
     s_{m,k}
     + \mathbf{n}_{m,k}^{\text{RSU}},
}\end{align} 
where {$\mathbf{y}^{\text{RSU}}_{m,k}$ is the RSU received signal on the $m$th subcarrier of the $k$th OFDM symbol, $\alpha^{\text{RSU}}$ is the complex sensing-channel gain incorporating two-way path loss and RCS of vehicle, $\tau^{\text{RSU}}$ is the round-trip delay of the sensing link, $f_{\text{D}}^{\text{RSU}}$} is the {Doppler frequency of the RSU received signal, and $\mathbf{n}_{m,k}^{\text{RSU}}$ is the sampled RSU noise}.

    \subsection{{Vehicle Received Signal Model for Communication}}
\label{sec: Vehicle Received Signal Model for Communication}

{The vehicular UE is equipped with $N_{\text{R}}^{\text{UE}}$ receive antennas. The UE receive beamforming direction is determined by} the two-step {predicted azimuth angle $\widetilde{\theta}_{t|t-2}$. The UE} receive steering vector {associated with the true azimuth angle $\theta_t$ is $\mathbf{b}_t^{\text{UE}} \triangleq \mathbf{b}(\theta_t, N_{\text{R}}^{\text{UE}})$, and the UE which uses maximum-ratio combining (MRC) beamforming has a receive beamforming vector $\mathbf{w}_{\text{R},t}^{\text{UE}} \triangleq \mathbf{b}(\widetilde{\theta}_{t|t-2}, N_{\text{R}}^{\text{UE}})$}.

 {Define ${G_{\text{R},t}^{\text{UE}}} \triangleq (\mathbf{w}_{\text{R},t}^{\text{UE},H}\mathbf{b}_t^{\text{UE}})^2$. The UE received signal is
}\begin{align}
{\mathbf{y}^{\text{UE}}_{m,k} \triangleq
\alpha^{\text{UE}}e^{j2\pi (f_{\text{D}}^{\text{UE}}kT - m\tau^{\text{UE}}\Delta f)}
\sqrt{G_{\text{R},t}^{\text{UE}}G_{\text{T},t}^{\text{RSU}}}
s_{m,k}
+ \mathbf{n}_{m,k}^{\text{UE}},
}\end{align} 
where {$\alpha^{\text{UE}}$ is the downlink channel gain, $\tau^{\text{UE}}$ is the downlink propagation delay, $f_{\text{D}}^{\text{UE}}$ is the downlink Doppler frequency, and $\mathbf{n}_{m,k}^{\text{UE}}$ is the sampled UE noise.
} 

 {With known attenuation, delay, and Doppler terms, the UE} signal can be {compensated as
}\begin{align}
 {\mathbf{y}^{\text{UE}}_{m,k} =\varrho^{-1}(\varrho\sqrt{G_{\text{R},t}^{\text{UE}}G_{\text{T},t}^{\text{RSU}}}
    s_{m,k} + \mathbf{n}_{m,k}^{\text{UE}}),
}\end{align} 
where {$\varrho \triangleq \alpha^{\text{UE}} e^{j2\pi (f_{\text{D}}^{\text{UE}}kT - m\tau^{\text{UE}}\Delta f)}$}.

    \subsection{{State Evolution and Measurement Model for RSU and UE}}
\label{sec: Evolution and Measurement Model}

 {The RSU extracts sensing measurements from the RSU received signal. After OFDM demodulation, the receiver in }\cite{jiamingReceiver} {obtains the time-frequency-domain channel state information (CSI). MUSIC-based angle estimation and two-dimensional FFT-based delay-Doppler estimation produce the raw sensing observation $[\widehat{\theta}_t,\widehat{\tau}^{\mathrm{RSU}}_t,\widehat{f}^{\mathrm{RSU}}_{\mathrm{D},t},\widehat{\alpha}^{\mathrm{RSU}}_t]$, where $\widehat{\tau}^{\mathrm{RSU}}_t$, $\widehat{f}^{\mathrm{RSU}}_{\mathrm{D},t}$, and $\widehat{\alpha}^{\mathrm{RSU}}_t$ denote the estimated round-trip delay, Doppler frequency, and sensing-channel gain, respectively. Since the sensing observations in this subsection are obtained at the RSU, the superscript $\mathrm{RSU}$ is omitted hereafter for notational simplicity unless otherwise specified. Considering} two-way {propagation, the Kalman measurement vector is constructed as $\mathbf{z}_t \triangleq [\widehat{\theta}_t,\frac{c\widehat{\tau}_t}{2},\frac{\lambda \widehat{f}_{\mathrm{D},t}}{2\cos{\widehat{\theta}_t}},\widehat{\alpha}_t]^{\mathrm{T}}$, where $c$ is the speed of light and $\lambda$ is the carrier wavelength. The true state vector is $\mathbf{x}_t \triangleq [\theta_t,d_t,v_t,\alpha_t]^{\mathrm{T}}$. After transforming the delay and Doppler estimates into distance and radial velocity, we adopt the following approximate additive measurement model: $\mathbf{z}_t=\mathbf{H}\mathbf{x}_t+\mathbf{n}_t$ with $\mathbf{H}=\mathbf{I}$, where $\mathbf{n}_t$ is the measurement noise vector. The prediction
  }\vspace{-0.2cm}
\begin{equation}
{{\left\{
\begin{aligned}
\widetilde{\theta}_{t|t-1} &= \widehat{\theta}_{t-1|t-1} + \frac{\widehat{v}_{t-1|t-1}}{\widehat{d}_{t-1|t-1}}  \Delta T \sin \widehat{\theta}_{t-1|t-1} + \omega_{\theta,t-1}, \\
\widetilde{d}_{t|t-1} &= \widehat{d}_{t-1|t-1} - \widehat{v}_{t-1|t-1} \Delta T \cos \widehat{\theta}_{t-1|t-1} + \omega_{d,t-1}, \\
\widetilde{v}_{t|t-1} &= \widehat{v}_{t-1|t-1} + \omega_{v,t-1}, \\
\widetilde{\alpha}_{t|t-1} &= \widehat{\alpha}_{t-1|t-1} \left( 1 + \frac{\widehat{v}_{t-1|t-1}}{\widehat{d}_{t-1|t-1}}  \Delta T \cos \widehat{\theta}_{t-1|t-1} \right)  + \omega_{\alpha,t-1},
\end{aligned}
\right.
\label{eq:state_evolution}
}}\end{equation}
{where $\widetilde{\theta}_{t|t-1}$, $\widetilde{d}_{t|t-1}$, $\widetilde{v}_{t|t-1}$, and $\widetilde{\alpha}_{t|t-1}$ denote the one-step predicted azimuth angle, propagation distance, velocity, and complex channel gain, respectively. The parameter $\Delta T$ is the prediction interval, and $\omega_{\theta,t-1}$, $\omega_{d,t-1}$, $\omega_{v,t-1}$, and $\omega_{\alpha,t-1}$ are the process noise terms}.

 {The two-step predicted UE receive beam direction is expressed as
}\begin{align}
\begin{split}
\color{blue}
&\widetilde{\theta}_{t|t-2} 
= 
\underbrace{
\widehat{\theta}_{t-2} 
+ \widehat{d}_{t-2}^{-1} \widehat{v}_{t-2} \Delta T \sin \widehat{\theta}_{t-2} 
+ \omega_{\theta,t-2}
}_{\text{first-step predicted angle}\widetilde{\theta}_{t-1|t-2}}
\\
& +
\underbrace{
\left(\widehat{d}_{t-2} - \widehat{v}_{t-2} \Delta T \cos \widehat{\theta}_{t-2} + \omega_{d,t-2}\right)
}_{\text{first-step predicted distance}\widetilde{d}_{t-1|t-2}}{}^{-1}
\underbrace{
\left(\widehat{v}_{t-2} + \omega_{v,t-2}\right)
}_{\widetilde{v}_{t-1|t-2}}
\Delta T
\\
& \times
\sin \left(
\underbrace{
\widehat{\theta}_{t-2} 
+ \widehat{d}_{t-2}^{-1} \widehat{v}_{t-2} \Delta T \sin \widehat{\theta}_{t-2} 
+ \omega_{\theta,t-2}
}_{\widetilde{\theta}_{t-1|t-2}}
\right)
+ \omega_{\theta,t-1}.
\end{split}
\label{eq: theta_prediction two step}
\end{align}

{After prediction and sensing, the Kalman filter corrects the predicted state using the measurement vector. With the first-order Taylor linearization in} \cite{FanKalman}, the {predicted state estimate and predicted covariance matrix are
}\begin{align}
{\widetilde{\mathbf{x}}_{t|t-1} }&{= \mathbf{F}_t\widehat{\mathbf{x}}_{t-1|t-1}, }\\
{\mathbf{P}_{t|t-1} }&{= \mathbf{F}_t\mathbf{P}_{t-1}\mathbf{F}_t^T + \mathbf{Q}_t,
}\end{align} 
where $\mathbf{F}_t$ is the state transition function{, $\widetilde{\mathbf{x}}_{t|t-1}$ is the predicted state estimate, $\mathbf{P}_{t|t-1}$ is the predicted error covariance matrix, $\mathbf{P}_{t-1}$ is the updated error covariance matrix at time instant $t-1$, and} $\mathbf{Q}_t$ {is} the process noise covariance matrix.

 The Kalman gain{, updated state estimate, and updated} covariance matrix are
   \begin{align}
{\label{eq: kalman gain}
}&{\mathbf{K}_t = {\mathbf{P}_{t|t-1}}({\mathbf{P}_{t|t-1} + \mathbf{R}_t})^{-1}, }\\
&{\widehat{\mathbf{x}}_{t|t} = \widetilde{\mathbf{x}}_{t|t-1} + \mathbf{K}_t(\mathbf{z}_t - \widetilde{\mathbf{x}}_{t|t-1}), }\\
&{\mathbf{P}_{t|t} = (\mathbf{I} - \mathbf{K}_t)\mathbf{P}_{t|t-1},
}\end{align} 
where $\mathbf{R}_t$ {is} the observation noise covariance matrix and $\mathbf{I}$ is the identity matrix.

The {Kalman filter recursively balances} estimation and prediction {uncertainties. Since vehicular motion can deviate from the assumed model due to} complex road conditions, the estimation update {is essential for} correcting the predicted state. {Accurate} estimation and prediction {further determine the beam steering accuracy}. Therefore, {the following section analyses the constellation-dependent} parameter estimation and angle prediction {RMSE}.

\section{Closed-form Analysis of Beamforming Gain and Error Propagation for Beam Misalignment}

 Based on the system model, this section {analyzes the dominant error sources in} the proposed sensing-assisted vehicular communication framework. {The received signal power is directly affected by the beamforming gain, which depends on the alignment between the actual and predicted beam directions. This establishes the error propagation chain from DDA estimation error to angular prediction error and finally to beamforming gain loss. Therefore}, we first {characterize} the beamforming gain {loss caused by angular misalignment, and then analyze the error propagation across the DDA estimation, one-step transmit beam prediction, and two-step receive beam prediction stages.
}

\begin{figure}
    \centering
    \includegraphics[width=0.7\linewidth]{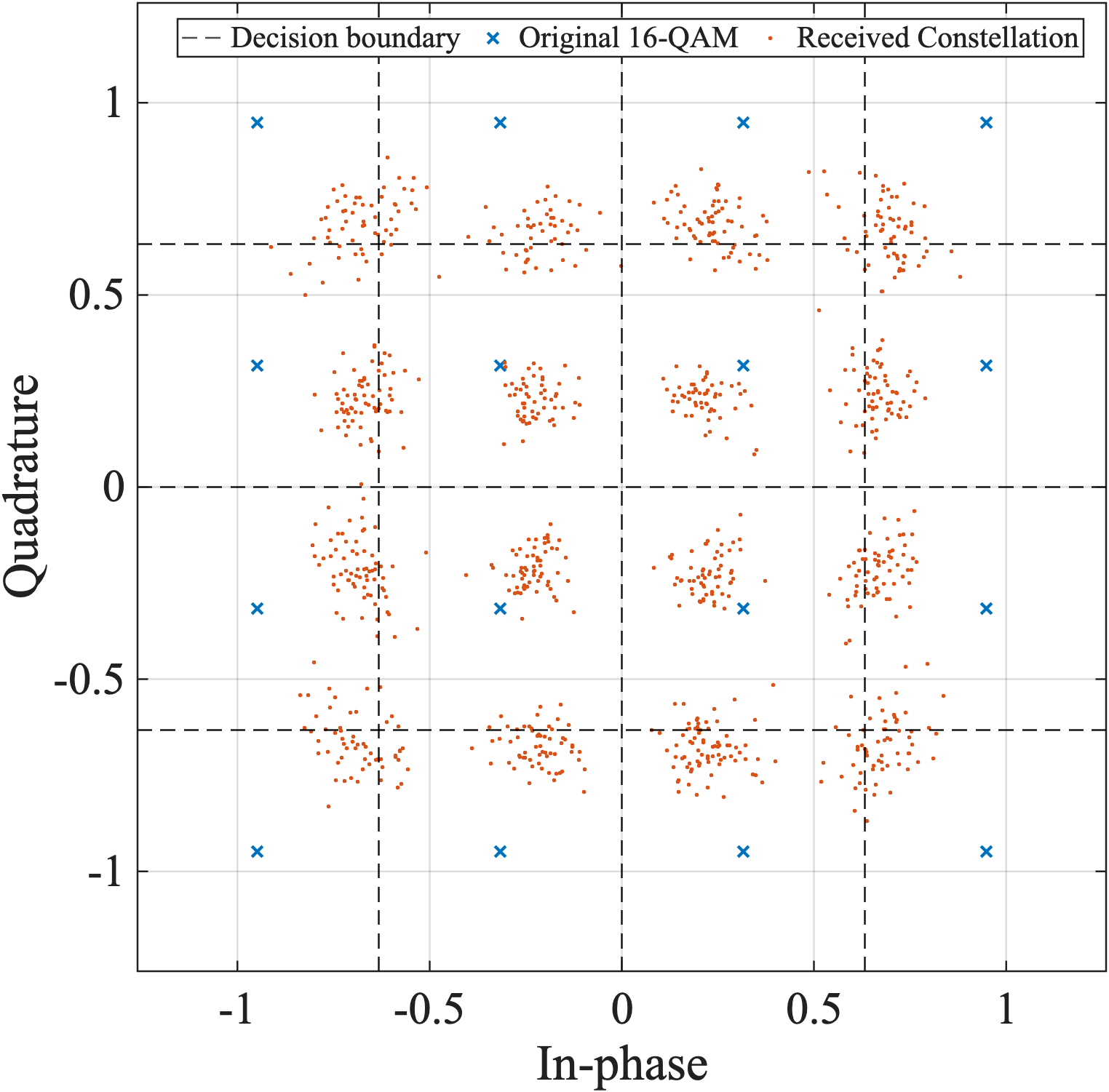}
\caption{Received {constellation under two-step prediction with $\mathrm{SNR}=20$ dB and $\rho_t=1$.}}
    \label{fig:constellation received}
\end{figure}

{The impact of such error propagation is illustrated in Fig.~\ref{fig:constellation received}, where the received 16-QAM constellation is distorted when $\mathbf{P}_{t|t-1} = \mathbf{R}_t$. As can be seen in the figure, the received constellation is seriously out of the decision boundary, and the observed constellation shrinkage indicates beamforming-induced energy loss, which seriously degrades constellation demapping and communication reliability}.

\subsection{{Beamforming Gain Loss under Angular Misalignment}}
    \label{sec: System Analysis - Beamforming Gain Loss}

 {We first characterize the beamforming gain degradation caused by angular misalignment. Let $\theta_t$ and $\widetilde{\theta}_t$} denote the true {and predicted azimuth angles at time instant $t$, respectively. For a ULA with $N$ antennas, the beamformer is $\mathbf{w}_t \triangleq \mathbf{a}(\widetilde{\theta}_t,N)$. The beamforming gain is
}\begin{align}
{\label{eq: org for BF gain}
    G_t }&{= | \mathbf{w}_t^H \mathbf{a}(  \theta_t,N)|^2
     = \frac{1}{N^2}\left|
\sum_{n=0}^{N-1}
 e^{-j\pi n
\left(
\cos\theta_t - \cos\widetilde{\theta}_t
\right)}
\right|^2 .
}\end{align} 

\begin{proposition}
\label{proposition: BF gain}
{Given the predicted azimuth angle $\widetilde{\theta}_t$ and the angular deviation $\Delta_{\theta_t} \triangleq \theta_t-\widetilde{\theta}_t$, the beamforming gain can be expressed as}
\begin{equation}
{\label{eq: Tx beamforming gain}
G_t(\Delta_{\theta_t},N)
=
\frac{1}{N^2}
\left|
\frac{
\sin\!\left(
N\pi
\sin\!\left(\widetilde{\theta}_t+\frac{\Delta_{\theta_t}}{2}\right)
\sin\!\left(\frac{\Delta_{\theta_t}}{2}\right)
\right)
}{
\sin\!\left(
\pi
\sin\!\left(\widetilde{\theta}_t+\frac{\Delta_{\theta_t}}{2}\right)
\sin\!\left(\frac{\Delta_{\theta_t}}{2}\right)
\right)
}
\right|^2.
}\end{equation}
\end{proposition}
 \textit{Proof}: {See Appendix A.
 }

\begin{figure}
    \centering
      \includegraphics[width=\linewidth]{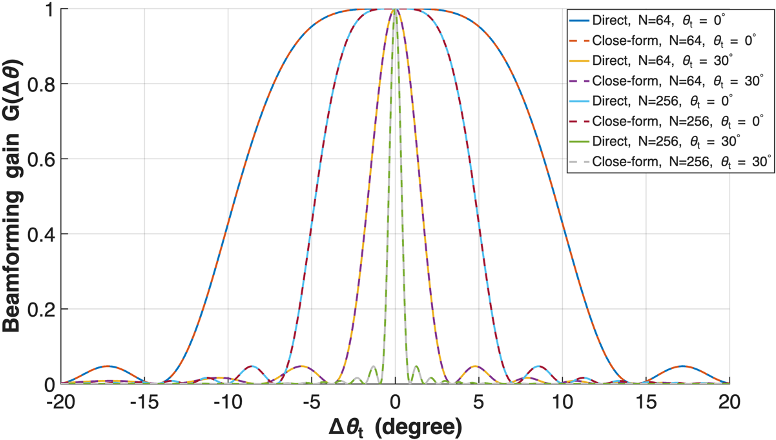}
     \caption{{Closed-form beamforming gain} under {angular mismatch}.}
    \label{fig:beam misalign - power}
\end{figure}

 {Proposition~\ref{proposition: BF gain} links angular prediction error to beamforming gain and applies to both} one-step and two-step {beam prediction.} Fig.~\ref{fig:beam misalign - power} {validates the closed-form expression against \eqref{eq: org for BF gain}. Larger arrays are more sensitive to angular errors. In addition}, for a fixed {angular deviation, the loss depends on} $\theta_t$ {and becomes more severe when the vehicle moves towards a more oblique direction}.

    \subsection{{DDA Parameter Estimation Error Analysis}}
\label{sec: System Analysis - Parameter Estimation}

 {This subsection analyzes the delay and Doppler estimation errors, which are mapped to range and radial-velocity errors for Kalman-based beam prediction. In the considered RSU-assisted communication system, a single vehicular UE is tracked for beamforming. To make the sensing analysis more general, we extend the DDA estimation model to a multi-target scenario and characterize the MF-} and {RF-based estimation errors. The estimated DDA parameters, i.e., delay, Doppler frequency, angle, and channel gain, are mapped to the Kalman-filter kinematic parameters, i.e., distance, velocity, angle, and channel gain. Hence, DDA estimation errors directly affect the kinematic prediction accuracy.
}

{We first analyze delay and Doppler estimation. In the considered} slow-time {regime, delay and Doppler estimation} can be decoupled, {so the range and velocity RMSEs are derived separately. Recall the received echo signal model in \eqref{eq: y_m,k}:
}\begin{align}
{     \mathbf{y}^{\text{RSU}}_{m,k} \triangleq
     \alpha^{\text{RSU}}
     e^{j2\pi \left(f_{\text{D}}^{\text{RSU}}kT - m\tau^{\text{RSU}} \Delta f\right)}
     \sqrt{G_{\text{R},t}^{\text{RSU}}G_{\text{T},t}^{\text{RSU}}}
     s_{m,k}
     + \mathbf{n}_{m,k}^{\text{RSU}},
}\end{align}

{To connect \eqref{eq: y_m,k} with the delay-estimation model, we first focus on the fast-time dimension by fixing the OFDM symbol index $k$. Since the Doppler-dependent phase term does not vary with the subcarrier index $m$, it can be absorbed into the effective complex sensing-channel gain. The beamforming gain is treated as part of the effective received-link gain in the subsequent RMSE analysis. Considering the multiple scatters in the surrounding environment for our single target, while stacking the received samples over the $M$ subcarriers gives a simplified expression for our received signal
}\begin{align}
 {\mathbf{y}_\tau = \boldsymbol{\alpha}^T\mathbf{T}\mathbf{X}_\tau + \mathbf{N}_\tau,
}\end{align} 
where {the attenuation vector $\boldsymbol{\alpha} = [\alpha_1,\alpha_2,\ldots,\alpha_J]^T \in \mathbb{C}^{J \times 1}$ contains the complex sensing-channel gains of $J$ targets, $\mathbf{T} = [\mathbf{t}(\tau_1),\mathbf{t}(\tau_2),\ldots,\mathbf{t}(\tau_J)]^T \in \mathbb{C}^{J \times M}$ is the delay steering matrix, and $\mathbf{X}_\tau=\mathrm{diag}(\mathbf{x}_\tau)$} is the diagonal {transmit-symbol matrix used for delay estimation}.

 {With matched filtering (MF),
}\begin{align}
{\mathbf{y}_{\tau,\mathrm{MF}} = \boldsymbol{\alpha}^T\mathbf{T}|\mathbf{X}_\tau|^2 + \mathbf{n}_{\tau,\mathrm{MF}} ,
}\end{align} 
 {and the delay estimate is obtained as
}\begin{align}
 {\widehat{\tau} = \arg\max_{\tau}|\mathbf{t}(\tau)^H\mathbf{y}_{\tau,\mathrm{MF}}^T|,
}\end{align} 
where {$\mathbf{t}(\tau) = [1,e^{-j2\pi \Delta f \tau},\ldots,e^{-j2\pi (M-1)\Delta f \tau}]^T$ is the} delay steering vector.

 {Similarly, for Doppler estimation, the slow-time received signal is
}\begin{align}
 {\mathbf{y}_{f_{\text{D}}} = \boldsymbol{\alpha}^T\mathbf{D}\mathbf{X}_{f_{\text{D}}} + \mathbf{N}_{f_{\text{D}}},
}\end{align} 
where {$\mathbf{D} = [\mathbf{d}(f_{\text{D},1}),\mathbf{d}(f_{\text{D},2}),\ldots,\mathbf{d}(f_{\text{D},J})]^T \in \mathbb{C}^{J \times K}$ is the Doppler steering matrix, and $\mathbf{X}_{f_{\text{D}}}=\mathrm{diag}(\mathbf{x}_{f_{\text{D}}})$} is the diagonal {transmit-symbol matrix used for Doppler estimation. After MF,
}\begin{align}
{\mathbf{y}_{f_{\text{D}},\mathrm{MF}}
}&{= \boldsymbol{\alpha}^T\mathbf{D}|\mathbf{X}_{f_{\text{D}}}|^2 + \mathbf{n}_{f_{\text{D}},\mathrm{MF}}, }\\
{\widehat{f}_{\text{D}}
}&{= \arg\max_{f_{\text{D}}} |\mathbf{d}(f_{\text{D}})^H\mathbf{y}_{f_{\text{D}},\mathrm{MF}}^T|.
}\end{align} 
where {$\mathbf{d}(f_{\text{D}}) = [1,e^{j2\pi T f_{\text{D}}},\ldots,e^{j2\pi (K-1)T f_{\text{D}}}]^T$ is the Doppler steering vector}.

 {Following} \cite{han2025constellationdesignofdmISACdata}, {let $\mu_4 \triangleq \frac{1}{L}\sum_{l=1}^{L}|s_l|^4$ denote the fourth moment of $\mathcal{S}=\{s_l\}_{l=1}^{L}$, and let $G_{\mathrm{MF}}$ denote the effective received-link gain for MF-based estimation}.

    \proposition{The RMSEs of range and velocity estimation using MF are given by
\label{proposition: MF RMSE}
\begin{align}
&\text{RMSE}_{r,z}^{\mathrm{MF}} \notag\\
&= \frac{\sqrt{3}c}{4\sqrt{2}\pi \Delta f\sqrt{M^3}}
\sqrt{\frac{(\mu_4 - 1)\sum_{j\neq z}^{J}|\alpha_j|^2 + \frac{\sigma_n^2}{G_{\mathrm{MF}}}}{|\alpha_z|^2}}, \\[8pt]
&\text{RMSE}_{v,z}^{\mathrm{MF}} \notag\\
&= \frac{\sqrt{3}c}{2\sqrt{2}\pi T f_c \cos\theta_z \sqrt{K^3}}
\sqrt{\frac{(\mu_4 - 1)\sum_{j\neq z}^{J}|\alpha_j|^2 + \frac{\sigma_n^2}{G_{\mathrm{MF}}}}{|\alpha_z|^2}}.
\end{align}
\textit{Proof}:} {See Appendix B.
} 

 {The MF-based RMSE depends on $\mu_4$. For} non-unimodular {constellations with $\mu_4>1$, scatterer interference affects the estimation accuracy, while this term vanishes for unimodular constellations with $\mu_4=1$}.

 element-wise division {by the transmitted symbols gives
}

\begin{align}
{\mathbf{y}_{\tau,\mathrm{RF}}  }&{= \boldsymbol{\alpha}^T\mathbf{T} + \mathbf{n}_{\tau,\mathrm{RF}},  }\\
{\mathbf{y}_{f_{\text{D}},\mathrm{RF}}  }&{= \boldsymbol{\alpha}^T\mathbf{D} + \mathbf{n}_{f_{\text{D}},\mathrm{RF}}.
}\end{align} 

 {Let $\nu_{-2} \triangleq \frac{1}{L}\sum_{l=1}^{L}|s_l|^{-2}$ denote} the inverse second moment of the {constellation, and let $G_{\mathrm{RF}}$ denote the effective received-link gain for RF-based estimation}.

    \proposition{The RMSEs of range and velocity estimation using RF are given by
\label{proposition: RF RMSE}
\begin{align}
\text{RMSE}_{r,z}^{\mathrm{RF}}
&= \frac{c\sqrt{3}\sigma_n}{2\sqrt{2}\pi \Delta f\sqrt{M^3}G_{\mathrm{RF}}}
\sqrt{\frac{\nu_{-2}}{|\alpha_z|^2}},\\
\text{RMSE}_{v,z}^{\mathrm{RF}}
&= \frac{c\sqrt{3}\sigma_n}{2\sqrt{2}\pi T f_c \cos\theta_z \sqrt{K^3}G_{\mathrm{RF}}}
\sqrt{\frac{\nu_{-2}}{|\alpha_z|^2}}.
\end{align}
\textit{Proof}:} {See Appendix B.
} 

 {Unlike MF, RF removes the scatterer-dependent interference term. Its estimation accuracy is mainly determined by the target channel gain, inverse second moment, noise power, and effective received-link gain}.

\begin{figure}
    \centering
    \includegraphics[width=0.9\linewidth]{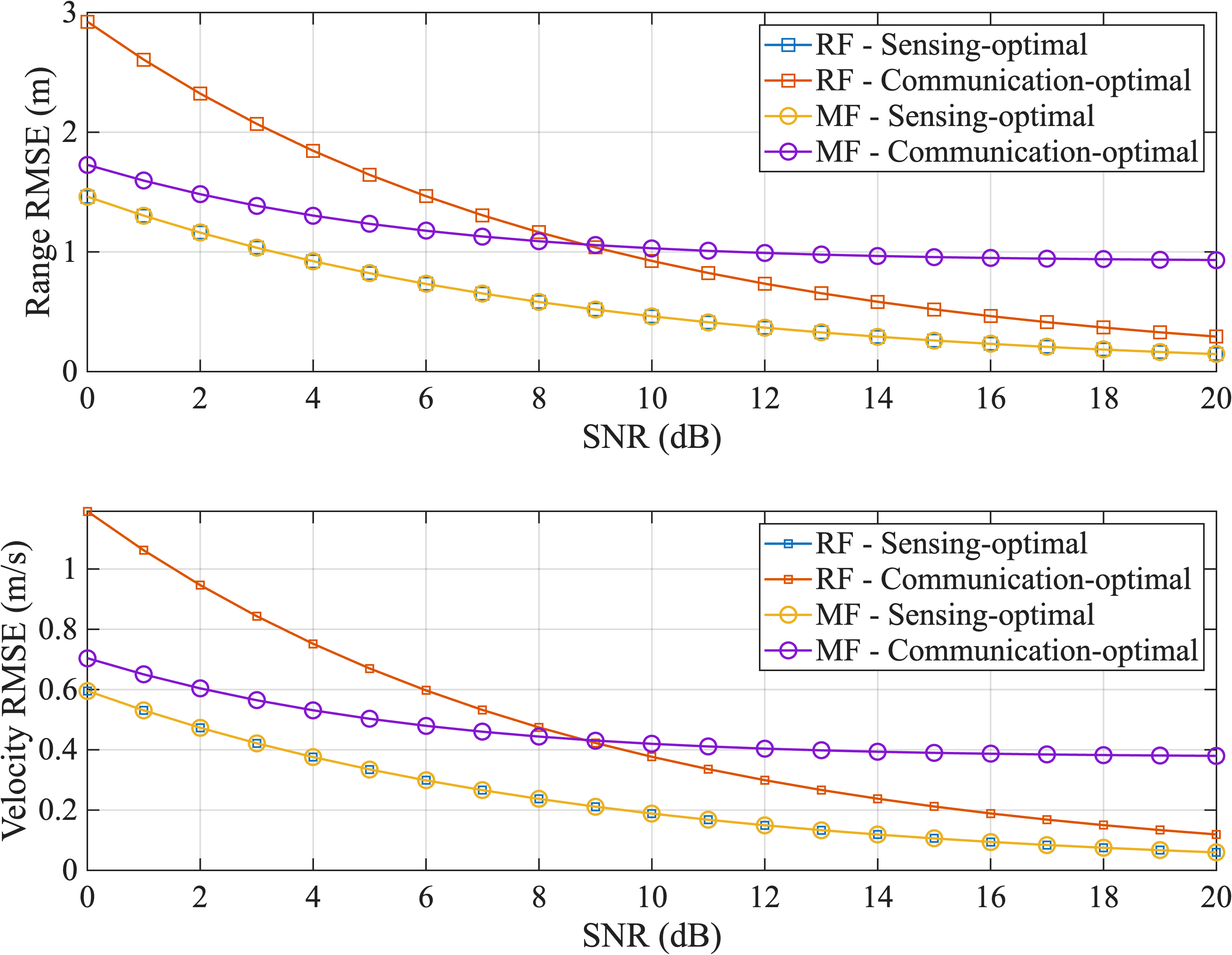}
    \caption{{Range} and velocity estimation {RMSE versus} SNR using MF and RF{.}}
    \label{fig:rmse_mf_rf_dv}
\end{figure}

 Fig.~\ref{fig:rmse_mf_rf_dv} {compares the range and velocity RMSEs of MF and RF under different constellations. With $10$ scatterers, MF shows a clear gap between communication-oriented constellations which gives the best communication performance and sensing-oriented constellations which are optimal for sensing, due to the fourth-moment term. For communication-oriented constellations, MF is preferable in} the {low-SNR regime, whereas RF becomes superior in the high-SNR regime}. For PSK constellations{, which are unimodular,} MF and RF {achieve nearly identical performance}. 

 \subsection{Transmit Beam Misalignment Analysis}
\label{sec: System Analysis - Transmit Beam Analysis}

{Having characterized the DDA parameter estimation errors, we analyze their impact on one-step transmit beam prediction. From the state evolution model in \eqref{eq:state_evolution}, the kinematic term in the one-step predicted azimuth angle is
} \begin{align}
    f(v_{t-1},d_{t-1}) = \frac{v_{t-1}}{d_{t-1}}\,\Delta T \sin(\theta_{t-1}).
\end{align}
 {Let $\delta v_{t-1}\triangleq \widehat{v}_{t-1}-v_{t-1}$ and $\delta d_{t-1}\triangleq \widehat{d}_{t-1}-d_{t-1}$ denote the velocity and distance estimation errors, respectively. By first-order Taylor expansion, the induced prediction-angle perturbation is
}\begin{align}
{\label{eq: delta theta}
\delta\theta_t \approx
\frac{\partial f}{\partial v_{t-1}}\delta v_{t-1} +
\frac{\partial f}{\partial d_{t-1}}\delta d_{t-1}.
}\end{align}  
 {The corresponding partial derivatives are
}\begin{align}
 {\label{eq: jacobian entry}
 \frac{\partial f}{\partial v} = \frac{\Delta T\sin(\theta_{t-1})}{d},
\qquad
\frac{\partial f}{\partial d} = -\frac{v\Delta T\sin(\theta_{t-1})}{d^2}.
}\end{align} 

 {Considering the angle process noise variance} $\sigma_{\omega_\theta}^2$,   the RMSE of the {one-step} predicted beam direction {is
}\begin{align}
{\epsilon_\theta  }&{=
  \sqrt{
\underbrace{
\left(\frac{\Delta T}{d}\sin\theta_{t-1}\right)^2 \epsilon_v^2
}_{\text{velocity term }T_v}
+
\underbrace{
\left(\frac{v\Delta T}{d^2}\sin\theta_{t-1}\right)^2 \epsilon_d^2
}_{\text{range term }T_d}
+
\underbrace{
\sigma_{\omega_\theta}^2
}_{T_n}
}.
\label{eq: rf final rmse transmit beam}
}\end{align}

\begin{figure}
    \centering
    \includegraphics[width=0.9\linewidth]{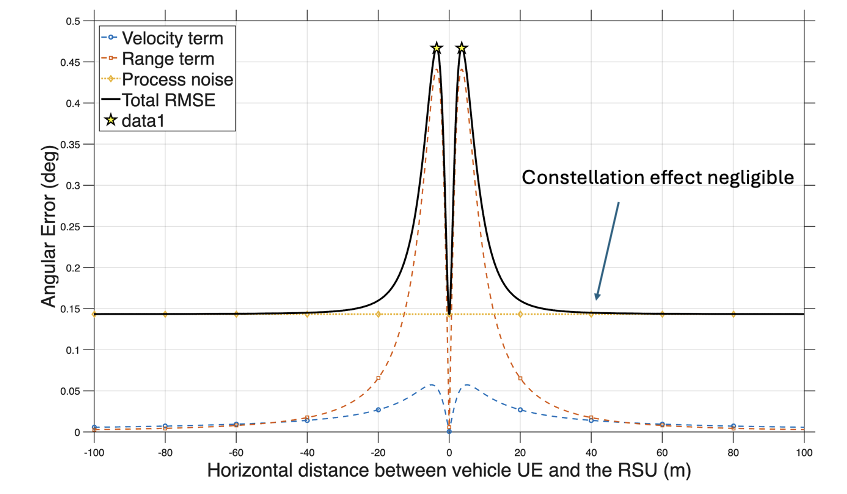}
    \caption{{Effect of range term, velocity term and process noise term on beam direction RMSE analysis w.r.t distance for RSU with height h = 5m}}
    \label{fig: rmse transmit beam_range}
\end{figure}
{Substituting the MF- and RF-based range and velocity RMSEs into \eqref{eq: rf final rmse transmit beam}, the one-step angle prediction RMSEs are obtained in \eqref{eq: expanded final mf theta rmse} and \eqref{eq: expanded final rf theta rmse}, respectively. According to \eqref{eq: Tx beamforming gain}, the corresponding transmit beamforming gains are $G_t(\epsilon_{1,\theta_{\mathrm{MF}}}, N_{\text{T}}^{\text{RSU}})$ and $G_t(\epsilon_{1,\theta_{\mathrm{RF}}}, N_{\text{T}}^{\text{RSU}})$. Further insight into the transmit beam prediction error can be obtained from Fig.~\ref{fig: rmse transmit beam_range}, which visualizes the three components in \eqref{eq: rf final rmse transmit beam}, namely the velocity-induced term $T_v$, the range-induced term $T_d$, and the process-noise term $T_n$, together with the overall beam prediction RMSE $\epsilon_\theta$}.

\begin{figure*}[ht]
    \begin{align}
{\epsilon_{1,\theta_{\mathrm{MF}}}}&{\approx
\sqrt{
\left(\frac{\Delta T}{d}\sin\theta_{t-1}\right)^2
\frac{3c^2\left((\mu_4 - 1)\sum_{j\neq z}^{J}|\alpha_j|^2 + \frac{\sigma_n^2}{G_{\mathrm{MF}}}\right)}{f_c^2 \cos^2\theta_z 8\pi^2 T^2 |\alpha_z|^2 K^3}
+
\left(\frac{v\Delta T}{d^2}\sin\theta_{t-1}\right)^2
\frac{3c^2\left((\mu_4 - 1)\sum_{j\neq z}^{J}|\alpha_j|^2 + \frac{\sigma_n^2}{G_{\mathrm{MF}}}\right)}{8\pi^2 (\Delta f)^2 |\alpha_z|^2 M^3}
+
\sigma_{\omega_\theta}^2
},
\label{eq: expanded final mf theta rmse}
}\\[0.4cm]
{\epsilon_{1,\theta_{\mathrm{RF}}}
}&{\approx
\sqrt{
\left(\frac{\Delta T}{d}\sin\theta_{t-1}\right)^2
\frac{3c^2\sigma_n^{2}\nu_{-2}}{f_c\cos\theta_z 8\pi^{2}T^{2}\lvert \alpha_{z}\rvert^{2}K^{3}G_{\mathrm{RF}}}
+
\left(\frac{v\Delta T}{d^2}\sin\theta_{t-1}\right)^2
\frac{3c^2\sigma_n^{2}\nu_{-2}}{8\pi^{2}\Delta f^{2}\lvert \alpha_{z}\rvert^{2}M^{3}G_{\mathrm{RF}}}
+
\sigma_{\omega_\theta}^2
}.
\label{eq: expanded final rf theta rmse}
}\end{align}  
  \hrule
  \end{figure*}

\begin{lemma} {The angular error exhibits a near-field peak and converges in the far field as the vehicle moves away from the RSU. 
  }

  \textit{Proof}: Denote the height of the RSU as $h$, and the horizontal distance between the RSU and the vehicle UE as $d_h$. Substituting $d(x)=\sqrt{d_h^2+h^2}$ and 
$\sin\theta(d_h)=\frac{d_h}{\sqrt{d_h^2+h^2}}$ into (34), the velocity-induced and range-induced terms can
be rewritten as
\begin{align}
T_v(d_h)
&=
\frac{\Delta T\,d_h}
     {d_h^2+h^2}
\epsilon_v \\
T_d(d_h)
&=
\frac{v\Delta T\,d_h}
     {(d_h^2+h^2)^{3/2}}
\epsilon_d.
\end{align}

At the closest-point position $d_h=0$, $T_v(0)=0,
T_d(0)=0$, which yields
\begin{equation}
\epsilon_\theta(0)
=
\sigma_{\omega_\theta}.
\end{equation}

As $|d_h|$ increases from zero, both
$\frac{|d_h|}{d_h^2+h^2}$
and
$\frac{|d_h|}{(d_h^2+h^2)^{3/2}}$
initially increase, leading to larger propagated
velocity and distance estimation errors.

For sufficiently large $|d_h|$,

\begin{equation}
T_v(d_h)
=
\mathcal{O}\!\left(\frac{1}{|d_h|}\right),
\qquad
T_d(d_h)
=
\mathcal{O}\!\left(\frac{1}{|d_h|^2}\right),
\end{equation}

and therefore

\begin{equation}
\lim_{|d_h|\rightarrow\infty}
T_v(d_h)=0,
\qquad
\lim_{|d_h|\rightarrow\infty}
T_d(d_h)=0.
\end{equation}

Substituting the above limits into (34) gives

\begin{equation}
\lim_{|d_h|\rightarrow\infty}
\epsilon_\theta(d_h)
=
\sigma_{\omega_\theta}.
\end{equation}

Since $\epsilon_\theta(d_h)$ is continuous, equals
$\sigma_{\omega_\theta}$ at $d_h=0$, increases due to the
propagated kinematic errors for finite $d_h$, and converges
back to the same limit as $|d_h|\rightarrow\infty$, it must
attain at least one local maximum at a finite road position.
The proof is completed.
\end{lemma}

{As} shown in Fig.~{\ref{fig: rmse transmit beam_range}, the velocity- and range-induced components vanish when the vehicle is directly underneath the RSU, since $\sin(\theta)=0$. As the vehicle moves away from this point, the geometric factor $\sin(\theta)$ activates the kinematic error-propagation terms, causing both $T_v$ and $T_d$ to increase. Among them, the range-induced component dominates over most of the considered trajectory because it is weighted by $v\epsilon_d$, whereas the velocity-induced component is weighted only by $\epsilon_v$. As the distance further increases, the geometric attenuation factors $1/d$ and $1/d^2$ suppress the kinematic error terms, and the total beam-direction RMSE gradually approaches the process-noise floor determined by $T_n$. This near-field peak and far-field convergence behavior is formalized in Lemma~1.
}

\begin{lemma}
{There exists a horizontal-distance region in which the
constellation effect becomes negligible.}

\textit{Proof}: Denote the RSU height by $h$ and the
horizontal distance between the vehicle and the RSU by $d_h$.
Using the same expression for $d(d_h)$ and $\sin{\theta(d_h)}$ as defined in Lemma ~1, the constellation-dependent contribution can be expressed as the combination of the range and velocity term,
\begin{equation}
T_c(d_h)
=
\left(
\frac{\Delta T d_h}{d_h^2+h^2}
\right)^2\epsilon_v^2
+
\left(
\frac{v\Delta T d_h}{(d_h^2+h^2)^{3/2}}
\right)^2\epsilon_d^2.
\end{equation}

For sufficiently large $|d_h|$,
\begin{equation}
T_c(d_h)
=
O\!\left(\frac{1}{d_h^2}\right)
+
O\!\left(\frac{1}{d_h^4}\right),
\end{equation}
and therefore $\lim_{|d_h|\rightarrow\infty}T_c(d_h)=0$.

Since $T_c(d_h)$ is continuous and monotonically decreases in
the far field, for any tolerance factor $\eta\in(0,1)$,
there exists a finite threshold $d_{h_\eta}>0$ satisfying $T_c(d_{h_\eta})=\eta\sigma_{\omega_\theta}^2$.

Hence, for all $|d_h|>d_{h_\eta}$, we have $T_c(d_h)\le
\eta\sigma_{\omega_\theta}^2$. Therefore, the beam prediction RMSE is dominated by the
process-noise term and the effect of constellation design
becomes negligible.
\end{lemma}
This completes the proof. Under the simulation setup used in Fig.~\ref{fig: rmse transmit beam_range}, setting $\eta=10^{-3}$ gives a critical distance of approximately $d_{h_\eta}=40$ m. This is consistent with Fig.~\ref{fig: rmse transmit beam_range}, where the range- and velocity-induced components become much smaller than the process-noise floor beyond this distance. Hence, after around $40$ m, the constellation-dependent estimation errors contribute less than $0.1\%$ of the process-noise power, and the overall one-step beam prediction RMSE is mainly determined by the process noise. Having achieved the above-mentioned discoveries, we next extend the analysis from one-step transmit beam prediction to the more challenging two-step receive beam prediction at the UE.

\subsection{{Receive Beam Misalignment Analysis}}
\label{sec: System Analysis - Receive Beam Analysis}

{Different from one-step transmit beam prediction, UE receive beamforming at time instant $t$ relies on } two-step {prediction from the Kalman-filtered state at time instant } $t-2${. Since no correction is available at the intermediate time instant $t-1$, the prediction errors are further propagated to time instant $t$, leading to more severe receive beam misalignment.
}

{To} avoid unnecessary coupling caused by {continuous Kalman iterations, we assume that $\widehat{\theta}_{t-2}\approx\theta_{t-2}$, while $\widehat{d}_{t-2}$ and $\widehat{v}_{t-2}$ contain constellation-dependent estimation errors.} From the state evolution model, substituting the intermediate {predicted states yields the nonlinear mapping
}\begin{equation}
{\widetilde{\theta}_{t|t-2} = f(\widehat{v}_{t-2}, \widehat{d}_{t-2}) + \eta_\theta,
}\end{equation} 
where $\eta_\theta$ collects {the} process noise terms {propagated from $\theta$, $d$, and $v$}.

Let {$\delta v\triangleq \widehat{v}_{t-2}-v_{t-2}$ and $\delta d\triangleq \widehat{d}_{t-2}-d_{t-2}$ denote the velocity and distance estimation errors}, respectively. A first-order Taylor expansion {around the true values} gives
\begin{equation}
\delta\theta_t
\approx
\frac{\partial \theta_t}{\partial v_{t-2}}\,\delta v
+
\frac{\partial \theta_t}{\partial d_{t-2}}\,\delta d
+
\eta_\theta.
\label{eq:theta_two_step_taylor}
\end{equation}
 {The partial derivatives are
} \begin{align}
\frac{\partial \theta_t}{\partial v_{t-2}}
&=
\frac{\Delta T}{d_{t-2}}\sin\theta_{t-2}
+
\frac{\Delta T}{d_{t-1}}\sin\theta_{t-1},
\label{eq:dtheta_dv_two_step}
\\
\frac{\partial \theta_t}{\partial d_{t-2}}
&=
-\frac{v_{t-2}\Delta T}{d_{t-2}^2}\sin\theta_{t-2}
-
\frac{v_{t-1}\Delta T}{d_{t-1}^2}\sin\theta_{t-1}.
\label{eq:dtheta_dd_two_step}
\end{align}

Assuming {the process noise variance of each kinematic parameter is time-invariant}, the effective noise term propagated to time $t$ is
   \begin{equation}
  \begin{split}
\eta_\theta &=
2\omega_{\theta}
+
\frac{\Delta T}{d_{t-1}}\sin\theta_{t-1}\,\omega_v
-
\frac{v_{t-1}\Delta T}{d_{t-1}^2}\sin\theta_{t-1}\,\omega_d.
\end{split} 
\end{equation} 
 {With mutually independent} zero-mean {noise terms}, its variance is
  \begin{align}
&{\mathrm{Var}(\eta_\theta)
=
 4\sigma^2_{\omega_{\theta}}
\nonumber}\\
&{+
\left(
\frac{\Delta T}{d_{t-1}}\sin\theta_{t-1}
\right)^2
\sigma^2_{\omega_v}
+
\left(
\frac{v_{t-1}\Delta T}{d_{t-1}^2}\sin\theta_{t-1}
\right)^2
\sigma^2_{\omega_d}.
}\end{align} 

Assuming $\delta v$ and $\delta d$ are uncorrelated with variances $\epsilon_v^2$ and $\epsilon_d^2$, respectively, the two-step predicted receive angle error variance is
\begin{align}
\epsilon_{\theta,n}^2
=
&\;
\left(
\frac{\partial \theta_t}{\partial v_{t-2}}
\right)^2
\epsilon_v^2
+
\left(
\frac{\partial \theta_t}{\partial d_{t-2}}
\right)^2
\epsilon_d^2
+
\mathrm{Var}(\eta_\theta).
\end{align}
  Accordingly, the RMSE of {the two-step predicted angle} is
\begin{equation}
\epsilon_{\theta,n}
=
\sqrt{
\left(
\frac{\partial \theta_t}{\partial v_{t-2}}
\right)^2
\epsilon_v^2
+
\left(
\frac{\partial \theta_t}{\partial d_{t-2}}
\right)^2
\epsilon_d^2
+
\mathrm{Var}(\eta_\theta)
}.
\label{eq:two step prediction final eq}
\end{equation}

Substituting {\eqref{eq:dtheta_dv_two_step}, \eqref{eq:dtheta_dd_two_step}}, and the {closed-form range and velocity estimation errors into \eqref{eq:two step prediction final eq}, the final two-step angle prediction RMSEs are obtained in \eqref{eq: expanded_final_mf_theta_rmse_second} and \eqref{eq: expanded_final_rf_theta_rmse_second}. According to \eqref{eq: Tx beamforming gain}, the corresponding receive beamforming gains are $G_t(\epsilon_{2,\theta_{\mathrm{MF}}}, N_{\text{R}}^{\text{UE}})$ and $G_t(\epsilon_{2,\theta_{\mathrm{RF}}}, N_{\text{R}}^{\text{UE}})$}.

\begin{remark}
 {Even with a perfect angle estimate} in the previous time block, angle prediction is still affected by {constellation-dependent range and velocity estimation errors, through either the fourth moment in MF or} the inverse second moment {in RF}. Therefore, when {$G_{\mathrm{MF}}=1$ and $G_{\mathrm{RF}}=1$, \eqref{eq: expanded final mf theta rmse} and \eqref{eq: expanded final rf theta rmse} provide lower bounds for one-step} prediction, while {\eqref{eq: expanded_final_mf_theta_rmse_second} and \eqref{eq: expanded_final_rf_theta_rmse_second} provide lower bounds for two-step prediction.
}

 \end{remark}

  \begin{figure*}[ht]
  \begin{equation}
{\epsilon_{2,\theta_{\mathrm{MF}}} \approx
  }{{\sqrt{
\begin{aligned}
\left(\frac{\Delta T}{d_{t-2}}\sin\theta_{t-2}
+
\frac{\Delta T}{d_{t-1}}\sin\theta_{t-1}\right)^2
\frac{3c^2\left((\mu_4 - 1)\sum_{j\neq z}^{J}|\alpha_j|^2 + \frac{\sigma_n^2}{G_{\mathrm{MF}}}\right)}{f_c^2 \cos^2\theta_z \, 8\pi^2 T^2 |\alpha_z|^2 K^3}
\quad+
\left(\frac{v_{t-2}\Delta T}{d_{t-2}^2}\sin\theta_{t-2}
+
\frac{v_{t-1}\Delta T}{d_{t-1}^2}\sin\theta_{t-1}\right)^2 \cdot \\
\frac{3c^2\left((\mu_4 - 1)\sum_{j\neq z}^{J}|\alpha_j|^2 + \frac{\sigma_n^2}{G_{\mathrm{MF}}}\right)}{8\pi^2 (\Delta f)^2 |\alpha_z|^2 M^3}
+2\sigma^2_{\omega_{\theta}}
+
\left(
\frac{\Delta T}{d_{t-1}}\sin\theta_{t-1}
\right)^2
\sigma^2_{\omega_v}
+
\left(
\frac{v_{t-1}\Delta T}{d_{t-1}^2}\sin\theta_{t-1}
\right)^2
\sigma^2_{\omega_d}
\end{aligned}
}
}} {\label{eq: expanded_final_mf_theta_rmse_second}
}\end{equation} 
\end{figure*}

\begin{figure*}[ht]
  \begin{equation}
{\epsilon_{2,\theta_{\mathrm{RF}}}\approx
  }{{\sqrt{
\begin{aligned}
\left(\frac{\Delta T}{d_{t-2}}\sin\theta_{t-2}
+
\frac{\Delta T}{d_{t-1}}\sin\theta_{t-1}\right)^2
\frac{3c^2\sigma_n^{2}\nu_{-2}}{f_c\cos\theta_z 8\pi^{2}T^{2}\lvert \alpha_{z}\rvert^{2}K^{3}G_{\mathrm{RF}}}
\quad+
\left(\frac{v_{t-2}\Delta T}{d_{t-2}^2}\sin\theta_{t-2}
+
\frac{v_{t-1}\Delta T}{d_{t-1}^2}\sin\theta_{t-1}\right)^2 \cdot \\
\frac{3c^2\sigma_n^{2}\nu_{-2}}{8\pi^{2}\Delta f^{2}\lvert \alpha_{z}\rvert^{2}M^{3}G_{\mathrm{RF}}}
+2\sigma^2_{\omega_{\theta}}
+
\left(
\frac{\Delta T}{d_{t-1}}\sin\theta_{t-1}
\right)^2
\sigma^2_{\omega_v}
+
\left(
\frac{v_{t-1}\Delta T}{d_{t-1}^2}\sin\theta_{t-1}
\right)^2
\sigma^2_{\omega_d}
\end{aligned}
}
}} {\label{eq: expanded_final_rf_theta_rmse_second}
}\end{equation} 
\end{figure*}

\section{System Performance Optimisation}
\vspace{-0.2cm}
Having obtained a numerical analysis of the whole system, we further propose two machine learning based optimisation approaches to improve the system performance. A learning based Kalman filter at the transmitter is first discussed, followed by a pre-compensation receiver. As applied in the previous discussion in Section~\ref{sec: System Analysis - Transmit Beam Analysis} that for a better prediction of beam direction, reducing the prediction noise variance is an important approach. In reality, state evolution described by \eqref{eq:state_evolution} is often partially unknown, highly nonlinear, or time-varying, making it difficult to construct precise analytical models required by extended Kalman filtering. By contrast, learning-based approaches leverage data-driven representations, such as neural networks, to implicitly capture complex and unknown sequential dynamics from observations. Similar to \cite{dahan2024uncertainty}, the forward process of our machine-learning based Kalman filter can be witnessed in Algorithm~\ref{alg: Deep Kalman Filtering}. As can be seen in the detailed algorithm, instead of using a fixed evolution model for prediction, we replace it with a DNN block, with the covariance matrix prediction block also replaced, {obtaining the estimated state vector $\widehat{\mathbf{x}}_t$}. Thus, with all blocks relating to the prediction replaced by the DNN and normally training with Kalman gain calculated, the MSE between the {true state and the estimated state} is used as the loss function \cite{dahan2024uncertainty}, given by $\mathcal{L} = \mathbb{E}\left[ | \mathbf{x}_t - \widehat{\mathbf{x}}_{t|t} |^2 \right]$.

Therefore, the unknown non-linear function for the trajectory can be learned in this way, therefore decreasing the prediction error while being deployed in the whole system. 
\begin{algorithm} 
    \caption{Deep Kalman Filtering} 
    \label{alg: Deep Kalman Filtering} 
    \begin{algorithmic}[1]
        \Statex \textbf{Input:} {estimated state $\widehat{\mathbf{x}}_{t-1}$ and covariance matrix $\mathbf{P}_{t-1}$
       }\Statex \textbf{Output:} {estimated state $\widehat{\mathbf{x}}_t$ and updated covariance matrix $\mathbf{P}_{t|t}$}.

  \For{$t=1$ \textbf{to} $T$}
    \State \textit{Prediction}:
    \State predict the current time position
    \State {$\widetilde{\mathbf{x}}_{t|t-1} \gets \mathbb{DNN}_{pos}(\widehat{\mathbf{x}}_{t-1})$
   }\vspace{0.1cm}
    \State predict the current time estimation covariance matrix
    \State $\mathbf{P}_{t|t-1} \gets \mathbb{DNN}_{cov}(\mathbf{P}_{t-1|t-1})$

    \vspace{0.2cm}
    \State \textit{Correction}:
    \State update the kalman gain accordingly
    \State $\mathbf{K}_t \gets {\mathbf{P}_{t|t-1}\mathbf{G}^T_{t}}({\mathbf{G}_t\mathbf{P}_{t|t-1}\mathbf{G}_t^T + \mathbf{R}_t})^{-1}$

    \vspace{0.1cm}
    \State update the current time estimation covariance matrix
    \State $\mathbf{P}_{t|t} \gets (\mathbf{I} - \mathbf{K}_t\mathbf{G}_t)\mathbf{P}_{t|t-1}$ 
    \vspace{0.1cm}
    \State update the current time position
    \State {$\widehat{\mathbf{x}}_{t|t} \gets \widetilde{\mathbf{x}}_{t|t-1} + \mathbf{K}_t(\mathbf{z}_t - \mathbf{G}_t\widetilde{\mathbf{x}}_{t|t-1})$
    \State compute loss function
    \State $\mathcal{L} = \mathbb{E}\left[ | \mathbf{x}_t - \widehat{\mathbf{x}}_{t|t} |^2 \right]$
    \State backward propagation and update Optimiser
 }\EndFor
    \end{algorithmic} 
\end{algorithm}

Moving forward to the receiver side, a training-based error compensation technique is proposed. As is widely known that the addition of some DNN blocks can improve the robustness of the network to specific noise conditions. Specifically in wireless communication scenarios, this technique has been used in \cite{9468646}, where unwanted complex noise exists in the received signal. Through a DNN block, the phase noise is shown to be well estimated and removed. Therefore, inspired by this kind of training technique, we add an extra DNN block for complex noise compensation prior to our Transformer-based decoder. In conventional wireless frameworks \cite{9468646}, off-line training for such network requires pilot overhead to explore the complex noise contaminated signals due to the unknown phase and amplitude noise variance. However, in our framework, with the aid of system geometry, we numerically derive the RMSE of the noise variances, so instead of relying on pilot overhead to detect the noise variance, we can do the training directly.

\section{Numerical Results and Discussions}
\label{sec: numerical results and discussion}

With the above-mentioned methodologies discussed, in this section, we first illustrate and explain the experimental set-ups in Section~\ref{sec: Results - experimental setup}. Then, following the set-ups, we then examine the effect of constellation on our system in Section~\ref{sec: Results - Constellation Design}. Finally in Section~\ref{sec: Results - Optimisation}, the experimental results for the two optimisation techniques are discussed.

\subsection{Experimental Setup}
\label{sec: Results - experimental setup}

As the road condition and actual vehicle movement are highly random, and as shown in \eqref{eq: kalman gain}, both the prediction uncertainty covariance matrix and the estimation error covariance matrix affect the final Kalman gain, to better explore the effect of constellation design on the sensing-assisted communication framework, we have the following experimental settings for error regimes, defined based on $\mathbf{P}_{t|t-1}$ and $\mathbf{R}_t$:
\begin{itemize}
\item High prediction uncertainty: $\rho_t \in (1.5, 3]$, where $\rho_t \triangleq \mathrm{tr}(\mathbf{P}_{t|t-1})/\mathrm{tr}(\mathbf{R}_t)$. In this regime, the prediction uncertainty dominates over the observation noise, resulting in a Kalman gain that places greater weight on the measurement update. 
\item Moderate prediction uncertainty: $\rho_t \in (0.5, 1.5]$. Here, the prediction and observation uncertainties are of comparable magnitude, and the Kalman filter combines prediction and measurement with similar weights.
\item Low prediction uncertainty: $\rho_t \in (0, 0.5]$. In this case, the observation noise dominates, yielding a smaller Kalman gain and a filter behaviour that relies primarily on the predicted state.
\end{itemize}

Moving forward to our receiver set-up, considering the propagation of noise for the two-step prediction which is discussed in details in Section~\ref{sec: System Analysis - Receive Beam Analysis}, the higher prediction noise variance is, the more inaccurate the two-step prediction is. However, in reality, the high prediction noise variance is commonplace as cars usually cannot move along a fixed direction. Therefore, it is not convincing to use the two-step prediction based receiver as the only receiver in our system. In the following experiments, two separate receivers are examined.
\begin{itemize}
    \item The vehicle demodulates by relying on the predicted parameters using the two-step prediction approach, denoted as prediction based
    \item The vehicle doesn't rely on the two-step predicted parameters, but uses MUSIC estimation as the DOA estimation, while using 2-D FFT for range and doppler estimation, denoted as estimation based.
\end{itemize}

\subsection{The Effect of Symbol Constellations}
\label{sec: Results - Constellation Design}

\begin{figure*}
    \centering
        \includegraphics[width=0.9\linewidth]{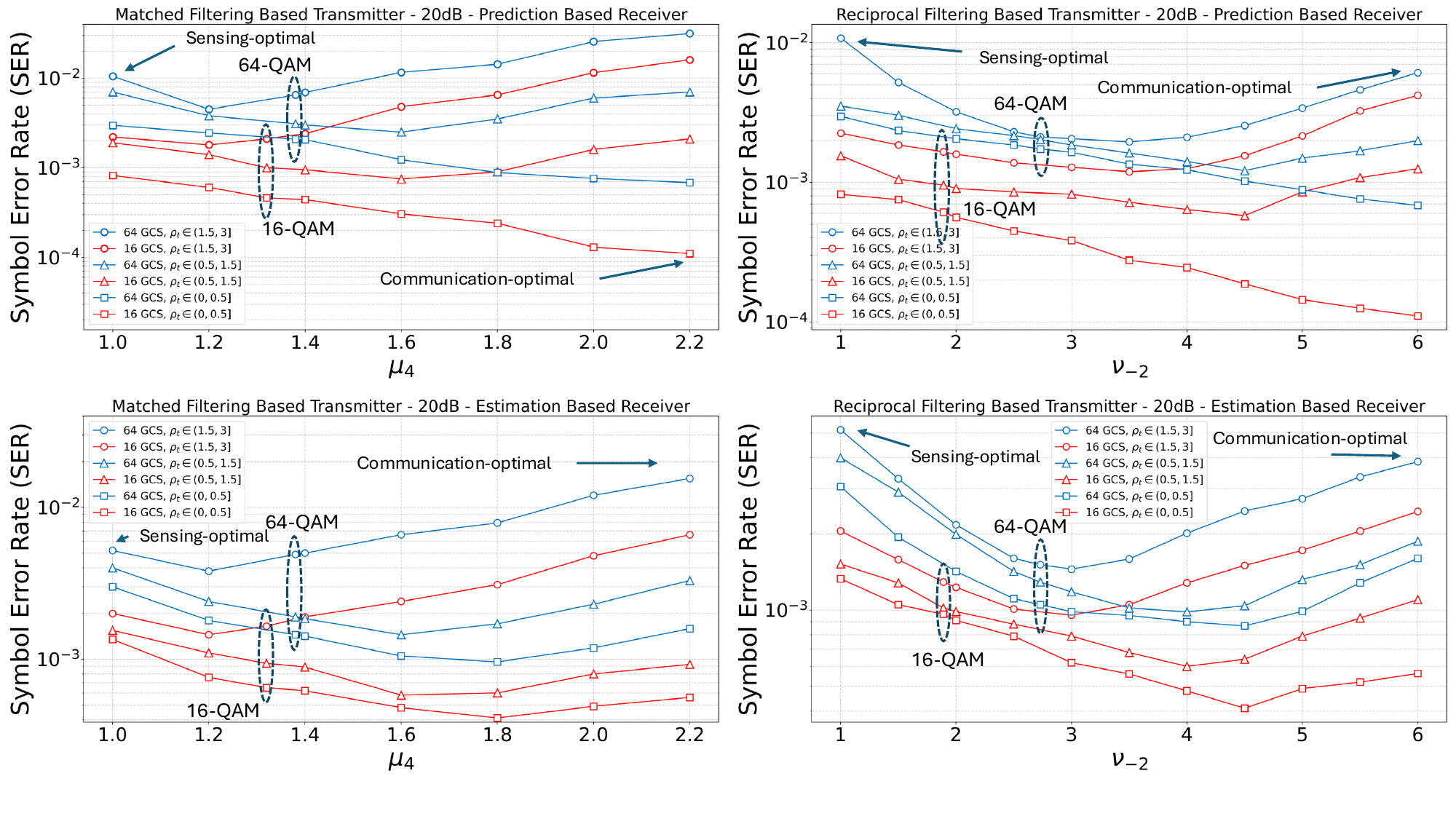}
      \caption{SER Plot for {estimation using matched filtering and reciprocal filtering when SNR = 20 dB}}
        \label{fig: MFRF SER plot}
  \end{figure*}

Briefly considering the prediction based sensing-assisted communication system, data detection at the vehicle relies on a sensing-informed channel model constructed from the predicted geometric parameters. Specifically, the receiver performs channel cancellation and subsequent demodulation using the predicted angle, range, and velocity as prior information. Consequently, constellation designs that yield different sensing accuracies, characterised by, e.g., the fourth-order moment (for MF) or the inverse second moment (for RF), lead to different errors in range–Doppler/DOA estimation, and hence induce different levels of residual interference after channel cancellation. A similar case happens in estimation based system, where constellation affects the estimation accuracy, therefore further affecting the channel cancellation and demodulating performance. 

To tackle the above problem and give an optimal solution, we numerically analyse the effect of constellation on the final communication performance of the whole system. Due to the deterministic-random trade-off, the more random the constellation is, the higher communication performance is. However, the communication-friendly constellations will lead to worse sensing performance, leading to further beam misalignment in our system and erroneous parameter estimation, therefore undermining the communication system, making the system unable to reach the ideal communication performance for the constellation. Therefore, in this subsection, such trade-off is further discussed. 

From the above estimation, it can be seen that for matched filtering estimation of parameters, the final result is highly affected by the clutter interference if non-unit modular constellation is used. In comparison, as can be seen in Fig.~\ref{fig:rmse_mf_rf_dv} that the clutter interference does not affect the final result while reciprocal filtering is used. Therefore, in this section, constellation analysis using both reciprocal filtering and matched filtering are analysed separately. To incorporate randomness in constellation generation, for both MF and RF, a total of 1000 independently generated geometric constellation shaping (GCS) realizations are considered, including modulation orders of 6 (64-GCS) and 4 (16-GCS), all sharing the same fourth-order moment. For each realization, the prediction and estimation noise variances are generated according to the aforementioned setup, followed by an end-to-end training procedure to obtain a learning-based decoder. The trained decoder is then evaluated under conditions with imperfect estimation and prediction. The final results are obtained by testing the trained decoder within this experimental framework.

\subsubsection{Matched {and Reciprocal} Filtering}

{The SER performances of both MF- and RF-based transmitters are shown in Fig.~\ref{fig: MFRF SER plot}. The left column corresponds to MF-based transmitters, while the right column corresponds to RF-based transmitters. The upper row presents the} prediction-based receiver, {where the UE demodulates the received signal using the two-step predicted geometric parameters, whereas the lower row shows the estimation-based receiver, where the receive-side parameters are directly estimated before demodulation. For MF, constellations with different fourth-order moments $\mu_4$ are tested, ranging from communication-optimal to sensing-optimal designs. For RF, constellations with different inverse second moments $\nu_{-2}$ are evaluated under the same setup, together with conventional 16-QAM and 64-QAM modulations.
}

{For MF-based systems, under} high prediction uncertainty, the SER exhibits a U-shaped behavior, with the minimum located closer to {sensing-oriented constellations}. This is {because matched filtering is} sensitive to the fourth-order moment $\mu_4$, {and non-unimodular constellations introduce additional clutter-dependent interference in range and Doppler estimation. In this regime,} sensing-oriented constellations {provide more accurate geometric information, thereby reducing beam misalignment and improving demodulation reliability. As the} prediction-noise {variance decreases, the sensing-induced error becomes less dominant, and the SER minimum gradually shifts toward} communication-oriented constellations{. When prediction becomes sufficiently accurate,} the {impact of constellation-induced estimation errors is weakened, and the SER is mainly} determined by the {intrinsic communication performance of the constellation. For} estimation-based {receivers, similar trends are observed, but} the performance gaps {among} different prediction-noise {regimes} are reduced. This is because {the receiver performs additional parameter estimation instead of relying entirely on the two-step predicted parameters. Hence, part of the accumulated prediction error can be corrected before demodulation. Nevertheless, constellation design still affects the RSU-side sensing accuracy and} transmit beam alignment, {so the sensing--communication trade-off remains visible in the SER curves.
} 

 {For RF-based systems, the dependence on constellation design becomes weaker than in the MF case. Since reciprocal filtering removes the clutter-interference term in parameter estimation, its sensing accuracy is mainly governed by $\nu_{-2}$, the noise power, and the received-link gain. Therefore, the SER gaps among different constellations are smaller, and the optimal operating point shifts more toward communication-oriented constellations. Under moderate prediction uncertainty, the SER still first decreases and then increases, indicating that an intermediate constellation design can better balance sensing accuracy and communication robustness.} Furthermore, under medium and high prediction-noise conditions, estimation-based {receivers generally outperform} prediction-based {receivers due to the error amplification introduced by} two-step prediction. In contrast, when prediction {accuracy is sufficiently high,} prediction-based {receivers can achieve better performance because they avoid additional receive-side estimation errors}. Overall, {Fig.~\ref{fig: MFRF SER plot} confirms that the optimal constellation depends jointly on the filtering method, receiver structure, and} prediction-noise {regime. It also verifies the main design insight of this paper: constellation shaping affects sensing accuracy, sensing accuracy determines beam prediction reliability, and beam prediction reliability ultimately governs the SER performance of the vehicular communication link}.

\subsection{Optimisation Techniques}
\label{sec: Results - Optimisation}

  \begin{figure}[!t]
      \centering
        \includegraphics[width=0.9\linewidth]{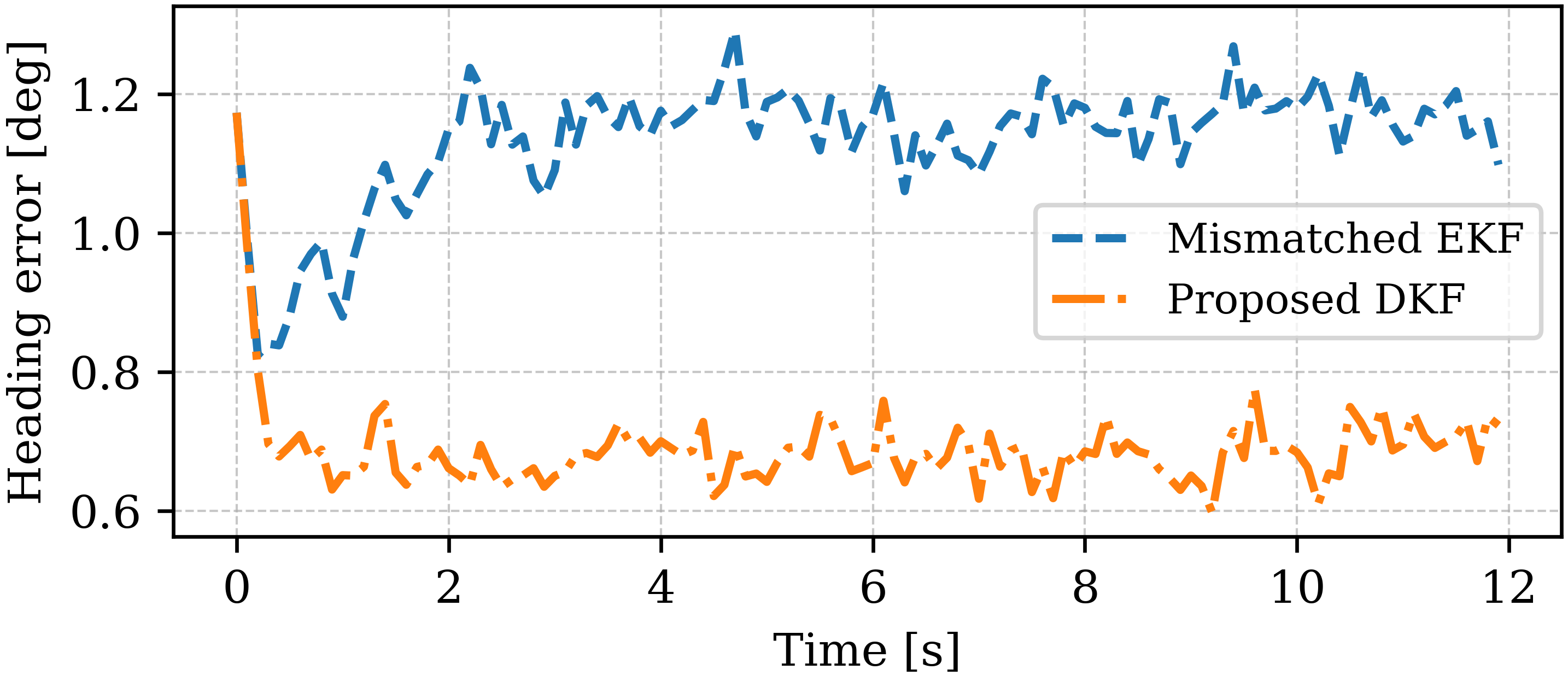}
      \caption{DKF Trajectory prediction performance{, $v=10$ m/s.}}
        \label{fig:trajectory_lkf}
  \end{figure}

Having discussed the SER performance for the system, we then conduct experiments about optimising the system. Starting from the Deep Kalman filter (DKF), we perform monte-carlo experiment over 1000 randomly generated non-straight trajectories, and compare our trained DKF with the EKF which still operates and predicts with straight line trajectory equations and the EKF with the precise evolution model. An experiment is performed while the actual trajectory prediction performance can be seen in Fig.~{\ref{fig:trajectory_lkf}}. As can be seen in the figure, the learnt DKF outperforms the erroneous EKF, and the unknown trajectory features can be well learned by the DNN block. The trajectory can be well predicted in this case, further improving the stability of this Kalman-assisted system. The detailed network parameters can be seen in Table. \ref{tab:deep_kalman_parameters}.

\begin{table}[t]
\centering
\caption{Deep-learning parameters of the proposed AI-assisted
Kalman filter, adopted from \cite{dahan2024uncertainty}.}
\label{tab:deep_kalman_parameters}
\renewcommand{\arraystretch}{1.08}
\begin{tabular}{lc}
\toprule
\textbf{Network Parameter} & \textbf{Setting} \\
\midrule
DNN architecture
    & FC--GRU--FC \\
Number of DNN blocks
    & $2$ \\
Number of GRU layers per block
    & $2$ \\
Input FC-layer width
    & $640$ \\
GRU hidden-state dimension
    & $64$ \\
Post-GRU FC-layer width
    & $64$ \\
Activation function
    & ReLU \\
Output dimension of $\mathrm{DNN}_{\mathrm{pos}}$
    & $4$ \\
Output dimension of $\mathrm{DNN}_{\mathrm{cov}}$
    & $16$ \\
Learning rate
    & $1\times10^{-3}$ \\
\bottomrule
\vspace{0.2cm}
\end{tabular}
\end{table}


  \begin{figure}[!t]
      \centering
        \includegraphics[width=0.9\linewidth]{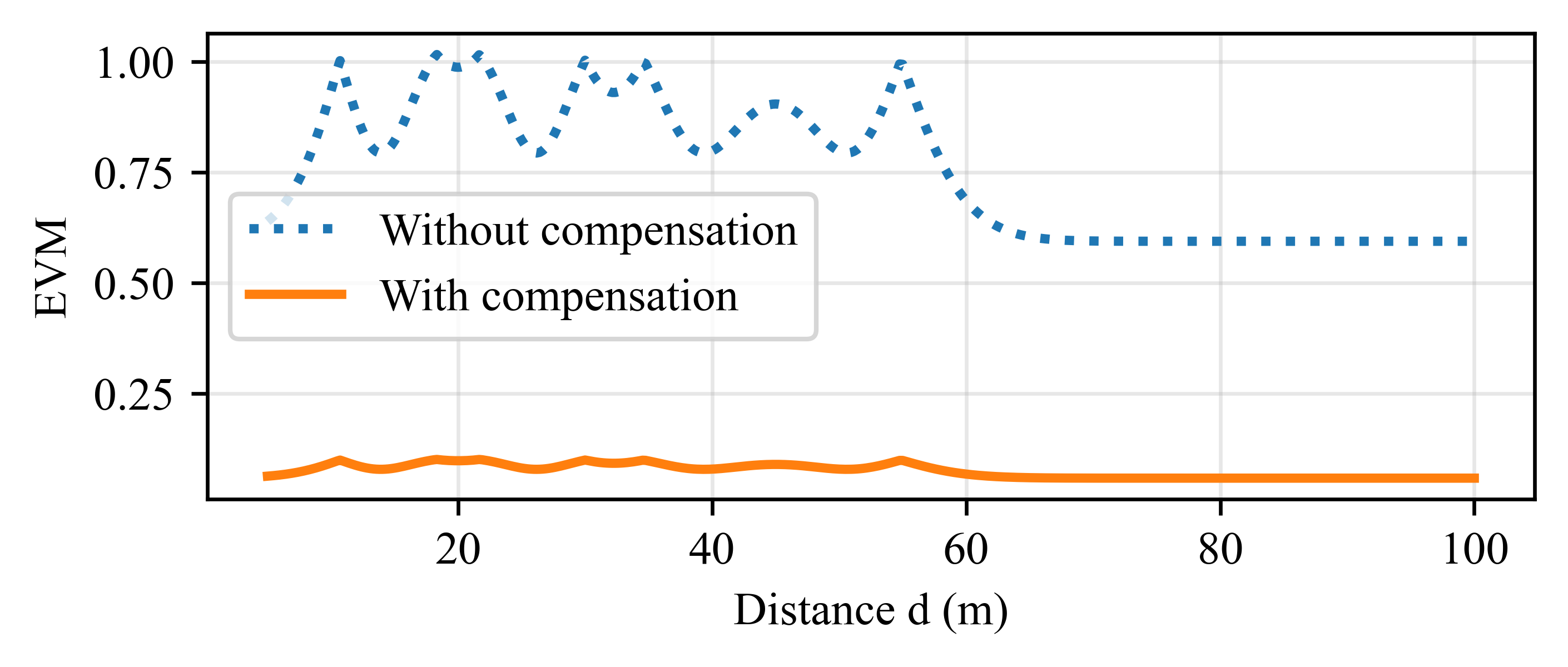}
      \caption{{EVM performance of the compensated model for $N_t=N_r=64$.}}
    \label{fig:evm_compensation}
    \vspace{-0.5cm}
\end{figure}

  \begin{figure}[t]
    \centering
    \includegraphics[width=\linewidth]{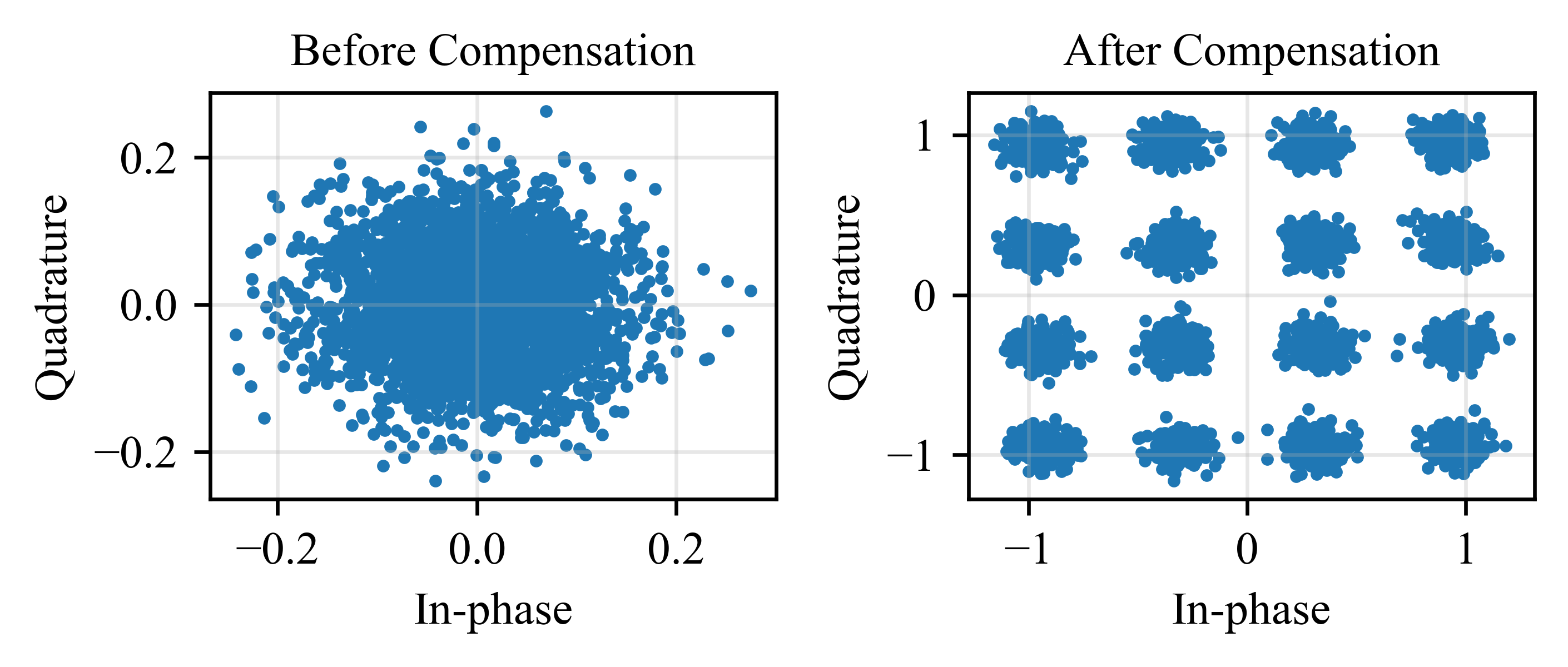}
    \caption{{Constellation comparison for compensated and original received constellation for $N_t=N_r=64$.}}
    \label{fig: constellation compensation}
    \vspace{-0.5cm}
\end{figure}
 {Moving forward to the receiver-side optimisation, a training-based pre-compensation strategy is proposed. Unlike the DKF framework, which aims to reduce prediction uncertainty, the proposed network directly compensates for the phase rotation and signal attenuation caused by beam misalignment. Specifically, the analytically derived angle-prediction RMSE is used to construct an equivalent distortion model, allowing the network to learn the corresponding compensation prior to symbol detection. However, as shown by \eqref{eq: expanded final mf theta rmse} and \eqref{eq: expanded final rf theta rmse}, the distortion level depends on both vehicle distance and velocity, making exact pre-training for all operating conditions difficult. To evaluate the proposed approach, we consider a scenario where the RSU-to-vehicle distance varies from 5 m to 100 m. Since MF and RF exhibit similar constellation-dependent trends, RF-based estimation at the transmitter and two-step prediction at the receiver are adopted as a representative example.}

{To evaluate the signal-level distortion caused by beam misalignment, we adopt the error vector magnitude (EVM) as the performance metric, which measures the normalized distance between the received constellation symbols and their ideal reference points, and therefore captures both amplitude attenuation and phase distortion in a unified manner. The resulting EVM performance} is shown in Fig.~{\ref{fig:evm_compensation}. Without compensation, beam-misalignment-induced phase distortion and beamforming-gain loss lead to consistently large EVM values across the considered distance range. After applying the proposed compensation network, the EVM is significantly reduced and remains relatively stable despite the continuously varying prediction-error statistics. This behaviour indicates that} the {network successfully captures the dominant distortion characteristics introduced by beam prediction errors and effectively restores the received signal quality over a wide range of operating conditions.
}

Further insight can be obtained from Fig.~\ref{fig: constellation compensation}, which compares the received constellations before and after compensation. Prior to compensation, the received symbols are severely dispersed around the origin, indicating substantial amplitude shrinkage and phase corruption caused by beam misalignment. Such distortion reduces the separation between neighbouring decision regions and increases the probability of symbol detection errors. After compensation, the constellation points become significantly more concentrated around the ideal 16-QAM locations, with clearly distinguishable symbol clusters. This result confirms that the proposed network can jointly compensate for both phase and amplitude impairments, thereby improving demodulation reliability and recovering a considerable portion of the communication performance lost due to sensing and prediction inaccuracies.

\section{Conclusions and Future Works}
 {This paper investigated} constellation design for sensing-assisted {vehicular communication systems under different prediction-noise regimes. A unified analytical framework was developed to characterize how constellation-dependent statistics affect sensing accuracy, beam prediction, and communication reliability. Specifically, closed-form expressions were derived for} the beamforming gain {under angular misalignment, the MF-} and {RF-based delay and Doppler estimation RMSEs, and the resulting one-step transmit and two-step receive beam prediction errors. The analysis revealed that the fourth-order moment $\mu_4$ governs MF-based sensing performance, while the inverse second moment $\nu_{-2}$ determines the corresponding RF-based behavior. Based on these results, the SER performance was evaluated for both prediction-based and estimation-based receivers, showing that the preferred constellation shifts from sensing-oriented to communication-oriented designs as prediction uncertainty decreases}. To further {enhance robustness,} a learning-based {Kalman filtering framework was proposed to reduce trajectory prediction errors under model mismatch, and a training-based} pre-compensation {strategy was developed to mitigate beam-misalignment-induced phase rotation and amplitude attenuation. Simulation results verified the derived analysis, demonstrated the effectiveness of both optimization strategies, and confirmed the critical coupling among constellation shaping, sensing accuracy, beam prediction reliability, and communication performance in dynamic vehicular ISAC systems}. As for future work, we will investigate more sophisticated vehicular scenarios, further researching the potential for constellation design in this framework.

\appendices
\section{{Proof of Proposition~\ref{proposition: BF gain}}}

To derive the final equation of the beamforming gain, starting from the initial definition in (17),
  \begin{equation}
{G_t
=
\left|
\mathbf{a}^H(\theta_{t},N)
\mathbf{a}(\widetilde{\theta}_t,N)
\right|^2 .
}\end{equation} 

For a uniform linear array with half-wavelength spacing, the steering vector is
\begin{equation}
\mathbf{a}(\theta,N)
=
\frac{1}{\sqrt{N}}
\left[
1,
e^{-j\pi\cos\theta},
\dots,
e^{-j\pi(N-1)\cos\theta}
\right]^T .
\end{equation}

Substituting the steering vector into the inner product gives
  \begin{equation}
{G_{t}
=
\frac{1}{N^2}
\left|
\sum_{n=0}^{N-1}
e^{-j\pi n(\cos\theta_{t}-\cos\widetilde{\theta}_t)}
\right|^2 .
}\end{equation} 

Using the trigonometric identity
\begin{equation}
\cos a-\cos b
=
-2\sin\frac{a+b}{2}\sin\frac{a-b}{2},
\end{equation}
  \begin{align}
{\cos\theta_{t}-\cos\widetilde{\theta}_t
=
-2\sin\bar{\theta}_{t}
\sin\left(\frac{\Delta\theta_{t}}{2}\right),
}\end{align} 
 {where $\Delta\theta_t \triangleq \theta_t-\widetilde{\theta}_t$ and $\bar{\theta}_t \triangleq (\theta_t+\widetilde{\theta}_t)/2$ denote the angular error and mean azimuth angle, respectively.
} 

Substituting into the exponential term yields
\begin{equation}
\sum_{n=0}^{N-1}
e^{j2\pi n
\sin\bar{\theta}_{t}
\sin\left(\frac{\Delta\theta_{t}}{2}\right)}.
\end{equation}

Using the identity $\left|
\sum_{n=0}^{N-1}e^{jnx}
\right|^2
=
\left(
\frac{\sin(Nx/2)}{\sin(x/2)}
\right)^2$,   \vspace{0.1cm}  the beamforming gain can be written in closed form as
  \begin{equation}
{G_t
=
\frac{1}{N^2}
\left|
\frac{
\sin\!\left(
N \pi
\sin\bar{\theta}_{t}
\sin\left(\frac{\Delta\theta_{t}}{2}\right)
\right)
}{
\sin\!\left(
\pi
\sin\bar{\theta}_{t}
\sin\left(\frac{\Delta\theta_{t}}{2}\right)
\right)
}
\right|^2.
\label{eq: appendix final step}
}\end{equation} 

As $\frac{\Delta_{\theta_{t}}}{2} << 1$, $\sin{\frac{\Delta_{\theta_{t}}}{2}} \simeq \frac{\Delta_{\theta_{t}}}{2}$. Therefore, substituting the approximation into \eqref{eq: appendix final step} completes the proof.

  \section{{Proof of Proposition~\ref{proposition: MF RMSE} \& Proposition~\ref{proposition: RF RMSE}}}
 \subsection*{\textit{Proof of Proposition~\ref{proposition: MF RMSE}}}
Starting from delay estimation, let us denote ${s}_{\text{mf}}(\tau) = \mathbf{t}(\tau)^H\mathbf{y}_{\tau, \text{mf}}^T$, and $f_{\text{mf}}(\tau) = |{s}_{\text{mf}}(\tau)|^2$. Therefore, for the source $z$ which is the desired target with its true time delay $\tau_z$, the MSE can be denoted by:
\begin{align}
    \text{MSE}_{r,z}^{\text{mf}} \approx \frac{\mathbb{E}[|\dot{f}_{\text{mf}}(\tau_z)|^2]}{|\mathbb{E}[|\ddot{f}_{\text{mf}}(\tau_z)|]|^2}.
    \label{eq: appendix MSE_init}
\end{align}
where $\dot{f}_{\text{mf}}(\tau_z)$ denotes the first-order derivative of $f_{\text{mf}}(\tau_z)$, and $\ddot{f}_{\text{mf}}(\tau_z)$ denotes the second-order derivative of $f_{\text{mf}}(\tau_z)$.

Therefore, denoting the first-order and second-order derivatives for ${s}_{\text{mf}}(\tau_z)$ as $\dot{s}_{\text{mf}}(\tau_z)$ and $\ddot{s}_{\text{mf}}(\tau_z)$, :
\begin{align}
    \dot{f}_{\text{mf}}(\tau_z) &= 2\Re\{\dot{s}_{\text{mf}}(\tau_z)s^*_{\text{mf}}(\tau_z)\} \approx 2N|\alpha_z|\Re\{\dot{s}_{\text{mf}}(\tau_z)\} \\
    \begin{split}
        \ddot{f}_{\text{mf}}(\tau_z)  &= 2\Re\{\ddot{s}_{\text{mf}}(\tau_z)s^*_{\text{mf}}(\tau_z) + \dot{s}_{\text{mf}}(\tau_z)\dot{s}^*_{\text{mf}}(\tau_z)\} \\
        &\approx 2N|\alpha_z|\Re\{\ddot{s}_{\text{mf}}(\tau_z)\}
    \end{split}
\end{align}

Therefore, we can express:
\begin{align}
    \mathbb{E}\{ |\dot{f}_{\text{mf}}(\tau_z)|^2\} = 2N^2|\alpha_z|^2\mathbb{E}[|\dot{s}_{\text{mf}}(\tau_z)|^2 ].\\
    |\mathbb{E}\{\ddot{f}_{\text{mf}}(\tau_z)\}|^2 = 4N^2|\alpha_z|^2\mathbb{E}[|\ddot{s}_{\text{mf}}(\tau_z)|^2 ].
\end{align}

Therefore, substituting these equations into \eqref{eq: appendix MSE_init} gives
\begin{align}
\label{eq: appendix MSE general final}
    \text{MSE}_{r,z}^{\text{mf}} = \frac{\mathbb{E}[|\dot{s}_{\text{mf}}(\tau_z)|^2]}{|\mathbb{E}[|\ddot{s}_{\text{mf}}(\tau_z)|]|^2}.
\end{align}

Expanding $s(\tau_z)$ and its derivatives into the exponential form, adding the MIMO beamformin gain, followed by element-wise multiplication, we can have: 
\begin{align}
    \mathbb{E}\{ |\dot{s}_{\text{mf}}(\tau_z)|^2\} &= 4\pi^2\Delta f^2\sum_{m=0}^{M-1}m^2\left((\mu_4 - 1)\sum_{j\neq z}^{K}|\alpha_j|^2 + \frac{\sigma_n^2}{G_{\text{mf}}}\right).\\
    |\mathbb{E}\{\ddot{s}_{\text{mf}}(\tau_z)\}|^2 &= -4\pi^2\Delta f^2\alpha_z\sum_{m=0}^{M-1}m^2.
\end{align}

Therefore, we have our final expression as:
\begin{align}
    \text{MSE}_{\tau, z}^{\text{mf}} &= \frac{3\left((\mu_4 - 1)\sum_{j\neq z}^{J}|\alpha_j|^2 + \frac{\sigma_n^2}{G_{\text{mf}}}\right)}{8\pi^2 (\Delta f)^2 |\alpha_z|^2 N^3}.
\end{align}
We have $\text{RMSE}_{\tau, z}^{\text{mf}} = \frac{c \ \sqrt{\text{MSE}_{r,z}^{\text{rf}}}}{2}$, which finishes the proof.

A similar proof is seen in the Doppler frequency estimation, where we represent $p_{\text{mf}}(f_D) = \mathbf{d}(f_D)^H\mathbf{y}_{f_D, \text{mf}}^T$. We then separately derive $\dot{p}_{\text{mf}}(f_{D_z})$ and $\dot{p}_{\text{mf}}(f_{D_z})$ through identical steps, with the dimension moving from the subcarrier $M$ to the symbol number $K$. Therefore, we have:
\begin{align}
    \text{MSE}_{f_D, z}^{\text{mf}} &= \frac{3\left((\mu_4 - 1)\sum_{j\neq z}^{J}|\alpha_j|^2 + \frac{\sigma_n^2}{G_{\text{mf}}}\right)}{8\pi^2 T_s^2 |\alpha_z|^2 K^3}.
\end{align}
Through $\text{RMSE}_{v, z}^{\text{mf}} = \frac{c * \sqrt{\text{MSE}^{\text{mf}}_{f_D}}}{f_c \cos\theta_z}$, we can finish our proof.

\subsection*{\textit{Proof of Proposition~\ref{proposition: RF RMSE}}}

Likewise, starting from the delay estimation, denote ${s}_{\text{rf}}(\tau) = \mathbf{d}(\tau)^H\mathbf{y}_{\tau, \text{rf}}^T$, following the process previously mentioned, we have:
\begin{align}
        \mathbb{E}\{ |\dot{s}_{\text{rf}}(\tau_z)|^2\} &= 4\pi^2\Delta f^2\sum_{m=0}^{M-1}m^2\frac{\sigma^2_n}{G_{\text{rf}}}\nu_{-2}.\\
    |\mathbb{E}\{\ddot{s}_{\text{rf}}(\tau_z)\}|^2 &= 16\pi^4\Delta f^4\alpha_z\left(\sum_{m=0}^{M-1}m^2\right)^2.
\end{align}

Following identical steps, we can achieve the two expectations for the Doppler estimation. Therefore, we have our MSEs represented as:
  \begin{align}
{\text{MSE}_{\tau,z}^{\text{rf}}
}&{=
\frac{3\sigma_n^2\nu_{-2}}{8\pi^{2}\Delta f^{2}\lvert \alpha_z\rvert^{2}N^{3}G_{\text{rf}}}, }\\
{\text{MSE}_{f_D,z}^{\text{rf}}
}&{=
\frac{3\sigma_n^{2}\nu_{-2}}{8\pi^{2}T_s^{2}\lvert \alpha_z\rvert^{2}K^{3}G_{\text{rf}}}.
}\end{align} 

Following $\text{RMSE}_{\tau, z}^{\text{rf}} = \frac{c \ \sqrt{\text{MSE}_{r,z}^{\text{rf}}}}{2}$, $\text{RMSE}_{v, z}^{\text{rf}} = \frac{c * \sqrt{\text{MSE}^{\text{rf}}_{f_D}}}{f_c \cos\theta_z}$, we can finish the proof.

\bibliographystyle{IEEEtran}
\bibliography{ref}

\end{document}